\documentclass[useAMS,usenatbib]{mnras}
\usepackage{amsmath}
\usepackage{appendix}
\usepackage{graphicx}
\usepackage{xfrac}
\usepackage{caption}
\usepackage{array}
\usepackage{longtable}
\usepackage{pdflscape}
\usepackage[figuresright]{rotating}
\usepackage{booktabs}
\usepackage[usenames, dvipsnames]{color}
\usepackage{subcaption}
\usepackage[flushleft]{threeparttable}
\usepackage[export]{adjustbox}
\usepackage{hyperref}
\allowdisplaybreaks

\usepackage{float}

\usepackage{media9}
\usepackage{animate}
\usepackage{etoolbox}

\emergencystretch=\maxdimen
\title[CSFH of bulges, disks, and spheroids]{Resolving cosmic star formation histories of present-day bulges, disks, and spheroids with ProFuse}
\author[Bellstedt et al.]
{Sabine Bellstedt,$^{1}$\thanks{Email: sabine.bellstedt@uwa.edu.au} Aaron S. G. Robotham,$^{1,2}$ Simon P. Driver,$^{1}$ Claudia del P. Lagos,$^{1,2}$ \newauthor Luke J. M. Davies,$^{1}$ and Robin H. W. Cook$^{1}$
\\
$^{1}$ ICRAR, The University of Western Australia, 7 Fairway, Crawley WA 6009, Australia\\
$^{2}$ ARC Centre of Excellence for All Sky Astrophysics in 3 Dimensions (ASTRO 3D)  \\
}

\begin{document}

\date{}

\pagerange{\pageref{firstpage}--\pageref{lastpage}} \pubyear{2023}

\maketitle

\label{firstpage}

\begin{abstract}
We present the first look at star formation histories of galaxy components using \textsc{ProFuse}, a new technique to model the 2D distribution of light across multiple wavelengths using simultaneous spectral and spatial fitting of purely imaging data. 
We present a number of methods to classify galaxies structurally/morphologically, showing the similarities and discrepancies between these schemes. 
We show the variation in component-wise mass functions that can occur simply due to the use of a different classification method, which is most dramatic in separating bulges and spheroids. 
Rather than identifying the best-performing scheme, we use the spread of classifications to quantify uncertainty in our results. 
We study the cosmic star formation history (CSFH), forensically derived using \textsc{ProFuse} with a sample of $\sim$7,000 galaxies from the Galaxy And Mass Assembly (GAMA) survey. Remarkably, the forensic CSFH recovered via both our method (\textsc{ProFuse}) and traditional SED fitting (\textsc{ProSpect}) are not only exactly consistent with each other over the past 8 Gyr, but also with the in-situ CSFH measured using \textsc{ProSpect}. 
Furthermore, we separate the CSFH by contributions from spheroids, bulges and disks. 
While the vast majority (70\%) of present-day star formation takes place in the disk population, we show that 50\% of the stars that formed at cosmic noon (8-12 Gyr ago) now reside in spheroids, and present-day bulges are composed of stars that were primarily formed in the very early Universe, with half their stars already formed $\sim$12 Gyr ago. 
\end{abstract}


\section{Introduction}

The extraordinary diversity of the present-day galaxy population is marked by a wide variety of galaxy properties. One such property is the physical structure of galaxies, in terms of galaxies that are disk-like in structure, spheroid-like, or the large population that contains multiple structural components. 

The build-up of structure in the very early Universe (prior to cosmic noon) is extremely chaotic, with the infall of gas causing immense star formation in large clumps, and lots of galaxy mergers making the definition of galaxy structure difficult \citep[as demonstrated by, for example,][]{lotz2004, elmegreen2005, lee2013}. By cosmic noon, however, structure is sufficiently well defined to describe galaxies in the context of substructures such as bulges and disks. As an example,  \citet{hashemizadeh2021} presented visual classifications of galaxies out to $z=1$, beyond which the fraction of structurally chaotic galaxies increases. 
From this point, there is an array of physical mechanisms that can result in the transformation and growth of structure with time. 
Disks can grow a bulge with time, either via disk instabilities funnelling disk material into a central bulge concentration (a structure often referred to as a pseudobulge, \citealt{kormendy2004}), or through mergers that add material straight into a bulge \citep[for example][]{barsanti2022}. Such a merger-origin bulge is frequently referred to as a ``classical" bulge. This two-phase mode is frequently implemented in semi-analytic models to grow bulges \citep[see for example][]{stevens2016, lagos2018a, husko2023}. 
Spheroidal galaxies can be formed from either disk or two-component systems via larger mergers that  destroy all existing structure \citep[as demonstrated using the Horizon-AGN simulations by][]{martin2018a}. 
Potentially, spheroidal systems could rebuild a disk through the accretion of gas, whose angular momentum creates star formation in a disk-like structure (predicted to occur in simulations by \citealt{steinmetz2002}, and evidence for which is potentially seen in studies by \citealt{moffett2012, fabricius2014}).
Finally, galaxies may experience no morphological change with time, with components instead growing through mechanisms like star formation (the case for disks), or via stellar mass build-up from mergers (usually  for spheroidal-like structures, as is the case for the mass growth in brightest cluster galaxies since $z\sim1-2$ \citealt{bellstedt2016, montenegro-taborda2023}). 

These galaxy structures/morphologies have been strongly linked to the star formation properties of the galaxies themselves. This was originally inferred very simply through a strong link between galaxy colour and its shape \citep[for example][who demonstrated that more concentrated galaxies had redder $u-r$ colours]{driver2006}, and then through measurements of the star formation rates as well \citep[like][]{lee2013}. By analysing the galaxy components themselves rather than simply characterising the overall galaxy shape, it was observed that the bulge fraction was also linked to the overall star formation \citep[see for example][]{fisher2009, guo2015}. 
This led to significant discussion and debate as to the impact of structural components like bulges on the overall star formation in galaxies \citep[for example][]{martig2009, bluck2014, cook2020}. 
Understanding the structural evolution of galaxies in the context of their star formation is therefore needed to build a more consistent picture of galaxy evolution. 

Pinpointing not only which evolutionary pathways have occurred across cosmic time, but also in what relative fractions, is an ongoing challenge in the field of galaxy evolution \citep[for example][]{casteels2014, robotham2014, bellstedt2017a, davies2022}. 
Making progress on this question observationally is hindered by a number of factors, the major one being that we are limited to observing individual galaxies in only a single snapshot. 
Given this limitation, two separate approaches must be taken to actually infer the temporal evolution of structure. 
The first of these is to compare galaxy populations across different epochs, to see how they are changing.
This approach is relied upon across the field of galaxy evolution to infer the evolution of most properties, including star formation rates \citep{driver2018, thorne2021}, metallicities \citep{ly2016, sanders2021}, velocity dispersion \citep{wisnioski2015}, velocity profile shapes \citep{tiley2019}, mass density profiles \citep{derkenne2021}, and a multitude of other properties, as well as the structure of galaxies \citep[as analysed by][]{hashemizadeh2021}.  
There are significant challenges in inferring evolution from different properties though, originating from progenitor bias \citep[just because populations at different epochs have similar mass distributions, does not mean one is the progenitor of the other, as commented by][]{dokkum1996, kaviraj2009}, selection effects, and observational limits. 
The other approach in inferring temporal evolution is to study the forensic histories of nearby galaxies themselves. 
What evidence is there in individual galaxies of evolving properties with time?

This two-pronged ``in-situ" versus ``forensic" approach has been well demonstrated in the pursuit of accurately constraining the cosmic star formation history (CSFH). 
The now-famous review by \citet{madau2014} brought together a suite of observational star formation rate density (SFRD) measurements across a wide redshift range to present a CSFH that showed the decline in star formation in the Universe since ``cosmic noon" at $z\sim2$. The exact nature of the CSFH prior to cosmic noon is still under debate, in particular because characterising the obscuration of star formation due to dust at high-$z$ is challenging \citep[for example][]{kistler2009, bourne2017, khusanova2020, bouwens2023, harikane2023}.
The alternate mechanism of deriving  the CSFH comes from a forensic analysis of a volume-complete sample of low-redshift galaxies. By deriving the star formation histories (SFH) of all galaxies within a volume of the Universe, the CSFH as a whole can be inferred. 
Stellar populations analysis techniques like SED modelling are usually employed to derive such star formation histories. To reliably construct the CSFH though, it is critical that any age-related degeneracies (such as the well-known age--metallicity degeneracy) are very carefully considered, so that no biases are introduced (see \citealt{conroy2013} for an overview of the general challenges faced when conducting SED fitting). In typical SED fitting implementations, metallicity is assumed to simply be constant over time, where at best this constant value is allowed to be free \citep[as employed in][]{leja2017, carnall2018a, iyer2019, johnson2021}, but at worst it is fixed to a single value \citep[typically solar, for example][]{yang2022, paspaliaris2023}. Such limitations make it challenging (or impossible) to accurately derive the CSFH in this manner, as the  derived CSFH peak is then offset from the observationally measured one due to the introduced biases \citep[for example][]{leja2019, carnall2019}. Recent implementations of more physical approaches to metallicity evolution in SED fitting in the code \textsc{ProSpect} \citep{robotham2020} have reduced these age--metallicity related biases, making it much more feasible to reliably reconstruct the CSFH from SED fitting  \citep[as demonstrated for the first time using SED fitting by][]{bellstedt2020b}. 

By applying this forensic-style approach not only to galaxies as a whole, but to their structural subcomponents, it is possible to study \textit{when} the stars in different galaxy structures formed. 
Actually isolating the stellar populations of galaxy components from imaging has historically required the successful completion of multiple steps. 
The first of these, is identifying the structural components of a galaxy through an analysis of the galaxy light profiles --- a process known as structural decomposition \citep[first presented in analysis such as][]{devaucouleurs1958}. 
Galaxy light can either be modelled as a single component \citep[usually with a S\'{e}rsic profile][]{sersic1963}, or with two or more components \citep[for example][]{cook2019}. 

Multi-wavelength imaging has also been used to increase the quality of galaxy decompositions, as shown by \citet{haussler2022}, using the \textsc{galapagos-2} code. 
Delving even deeper, two-dimensional decompositions can be extracted across multiple wavelengths to generate an SED per component. 
In a separate step, this can then be modelled with an SED-fitting code to extract forensic properties like star formation histories. 
Works that have used this approach include for example \citet{dimauro2018} who applied this to 17,600 galaxies from the CANDELS survey, to produce catalogues of bulge and disk properties.   
Because of this multi-step approach, such studies are complex and intricate, with potentially limited room for scientific interpretation. 

With the recent development of \textsc{ProFuse} \citep{robotham2022}, it is now possible to conduct image decomposition and SED fitting in a single, self-consistent and physically motivated step. 
This is quite distinct to similar approaches that have been developed recently to extract bulge and disk stellar populations from IFU data (see \citealt{robotham2022} for a more detailed discussion of these other techniques). 
In this work, we aim to present the first analysis applying this technique to a volume-limited sample, extracting the cosmic star formation history forensically for bulges, disks, and spheroids directly. 

The data used in our analysis are described in Sec. \ref{sec:Data}, with our \textsc{ProFuse} analysis technique and method described in Sec. \ref{sec:Method}. We present a detailed discussion of how our structural classifications compare to other methods in Sec. \ref{sec:ClassificationRegimeComparison}, and the results from our analysis are presented in Sec. \ref{sec:Results}. We discuss implications of our results in Sec. \ref{sec:Discussion}, and summarise our conclusions in Sec. \ref{sec:Conclusions}. 

The cosmology assumed throughout this paper is $H_0 = 67.8\,\rm{km}\,\rm{s}^{-1}\,\rm{Mpc}^{-1}$,  $\Omega_m = 0.308$ and $\Omega_{\Lambda} = 0.692$ \citep[consistent with a Planck 15 cosmology][]{planckcollaboration2016}. 

\section{Data}
\label{sec:Data}

This study utilises the wealth of data from the now-public Galaxy And Mass Assembly (GAMA) survey\footnote{\url{http://www.gama-survey.org/}} \citep{driver2011, liske2015, driver2022}, which is a galaxy redshift survey conducted on the Anglo Australian Telescope covering 250 square degrees, with a total of 303,542 redshifts \citep{driver2022}. 
We focus on a sample of 6,664 $z<0.06$ galaxies from the three equatorial regions of the GAMA survey (G09, G12, and G15) with high-quality redshift measurements, analysed in \citet{bellstedt2020b} and also \citet{robotham2022}.

We use multi-wavelength imaging from the GAMA DR4 \citep{driver2022}, as outlined by \citet{bellstedt2020a} in the $u$/$g$/$r$/$i$/$Z$/$Y$/$J$/$H$/$Ks$ bands. 
In the optical bands, the imaging originates from the VST \citep[VLT Survey Telescope, ][]{arnaboldi1998}, collected through the KiDS survey \citep[Kilo Degree Survey, ][]{dejong2013}, and in the infrared the imaging originated from VISTA \citep[Visible and Infrared Survey Telescope for Astronomy, ][]{emerson2006, dalton2006}, collected through VIKING \citep[VISTA Kilo-degree INfrared Galaxy survey, ][]{edge2013}. 
This imaging is all aligned using \textsc{SWarp} \citep{bertin2010} to a common pixel scale of 0.339 arcseconds (matching the native pixel scale of the VISTA image, for further details on this process see \citealt{bellstedt2020a}). 

Sources in our sample generally have at least 200 pixels of imaging data within the source segment (up to 50,000), demonstrating that in this redshift range we have sufficient 2D resolution elements for our spatial analysis. Futhermore, the imaging is sufficiently deep to ensure detections in all bands. As shown in figure 18 of \citet{bellstedt2020a}, the 5$\sigma$ surface brightness limit of the imaging data range between 22.5 to 25 mag depending on the band, which is substantially deeper than the $r<19.6$ galaxies in this sample. 

To conduct various completeness corrections related to the use of a volume-limited sample throughout this work, we make use of \textsc{ProSpect}-derived $z_{\rm max}$ values, which estimate the redshift to which any given galaxy is observable given the GAMA DR4 95\% completeness limit of $m_r < 19.65$ \citep{driver2022a}. 
These were derived by generating the best-fitting SED for each galaxy (as derived by the \textsc{ProSpect} fits from \citealt{bellstedt2020b} using the updated photometry presented by \citealt{bellstedt2020a}), and then regenerating this SED at a range of redshifts, identifying the value at which the magnitude limit is reached. 
Given the unmasked area of 169.29 square degrees covered by the sample in this work, the $z_{\rm max}$ value can then be converted to a $V_{\rm max}$, to represent the fraction of the observed volume within which the galaxy is observable.

\section{Method}
\label{sec:Method}

\subsection{\textsc{ProFuse}}

The tool that we use to conduct the simultaneous spectral and photometric decomposition of our sources is \textsc{ProFuse} \citep{robotham2022}. 
This combines the SED-fitting capabilities of \textsc{ProSpect} \citet{robotham2020}, the structural modelling capabilities of \textsc{ProFit} \citep{robotham2017}, and the source-finding capabilities of \textsc{ProFound} \citep{robotham2018}. 
While \textsc{ProFuse} as a tool in its entirety is relatively new, it has been developed over many years, through the gradual creation and application of its individual building blocks. 

The first of these tools was \textsc{ProFit} \citep{robotham2017}, which was developed to facilitate light profile fitting of galaxies in a generative, Bayesian manner. Previous codes had been prone to model error, making 2D light modelling of large samples of galaxies unwieldy, and the Bayesian nature of \textsc{ProFit} improved this behaviour. \textsc{ProFit} was wielded by studies such as \citet{cook2019} and \citet{cook2020} to assess the impact of galaxy bulges in xGASS \citep{catinella2010} on the scatter of HI scaling relations and the star forming main sequence,  and later by \citet{hashemizadeh2022} to study the evolution of bulges and disks through cosmic time in the DEVILS survey \citep{davies2018}. 

Separately to \textsc{ProFit}, there was a need to develop a generative and Bayesian tool that was capable of conducting SED fitting in a flexible and physically motivated manner. This inspired the development of \textsc{ProSpect} \citep{robotham2020}. With a particular desire to flexibly extract unbiased star formation histories (which in turn would ensure more accurate properties such as total stellar mass), \textsc{ProSpect} was applied to a volume-limited sample by \citet{bellstedt2020b}, where the capacity for this code to forensically reconstruct the cosmic star formation history was demonstrated. This was only possible due to the physically motivated implementation of metallicity evolution within \textsc{ProSpect}, shown to realistically evolve metals in galaxies as demonstrated by the correct inferred evolution of the mass--metallicity relation \citep{bellstedt2021}. This SED fitting tool was soon shown to work just as effectively at higher redshifts using the DEVILS survey by \citet{thorne2021}, with the link between metallicity and star formation histories explored by \citet{thorne2022a}. These studies demonstrated that \textsc{ProSpect} could self-consistently model the local Universe and infer the correct evolution of stellar mass, star formation rate and metallicity. This is as yet undemonstrated by other SED fitting codes, of which numerous exist with various degrees of similarity, e.g \textsc{Prospector} \citep{leja2017, johnson2021}, \textsc{BAGPIPES} \citep{carnall2018a},  \textsc{MAGPHYS} \citep{dacunha2008}, \textsc{CIGALE} \citep{noll2009, boquien2019}, and \textsc{BEAGLE} \citep{chevallard2016} (see \citealt{thorne2021} for a direct comparison of their utility and capabilities). This physically motivated implementation lends itself incredibly well to the incorporation of additional phenomena, as shown through the inclusion of an AGN component by \citet{thorne2022}, which was even extended to include radio data in the fitting process \citep{thorne2023}. Through application of this self-consistent stellar and AGN modelling, the link between star formation and AGN activity over 12.5 Gyr of cosmic time could be studied using GAMA and DEVILS by \citet{dsilva2023}. 

The final implement in the \textsc{ProFuse} toolbag is a source detection routine, capable of identifying and isolating the flux of the source object, but also nearby stars to facilitate a local extraction of the image point spread function (PSF). This source extraction software is \textsc{ProFound} \citep{robotham2018}, which has been readily used since its development to extract photometry of large, multi-wavelength imaging datasets for surveys \citep{bellstedt2020a, davies2021}.  

While \textsc{ProFuse} is described in great detail by \citet{robotham2022}, we describe our implementation of \textsc{ProFuse} in the following sections. 

\subsubsection{Structural models}

Separate structural models have been applied to each galaxy, similar to those used by \citet{robotham2022}. We generally use a S\'{e}rsic profile \citep{sersic1963} to model individual structural components in galaxies. The separate models are:

\begin{itemize}
	\item \textbf{BD}: \textbf{B}ulge+\textbf{D}isk mode, with an exponential disk component with S\'{e}rsic $n=1$ and a circular de Vaucouleurs bulge with $n=4$;
	\item  \textbf{FS}: \textbf{F}ree \textbf{S}\'{e}rsic mode, featuring a single component with a free S\'{e}rsic index; 
	\item  \textbf{PD}: \textbf{P}SF bulge + \textbf{D}isk mode, where the disk component has S\'{e}rsic $n=1$, however the bulge is modelled by a point source that is convolved with the image PSF (is different in each band). 
	\item  \textbf{DD}: \textbf{D}isk + \textbf{D}isk mode, where both disk components have S\'{e}rsic $n=1$, but the axial ratios and position angles of the disks are free. We expect that this is an infrequently preferred model, however may be appropriate in cases with very prominent bars or colour gradients. 
\end{itemize}

The DD run was not presented in \citet{robotham2022}, however as we will show it is the best-selected model for only a small number of sources. 

The implementation of a two-component model with a disk and a free-S\'{e}rsic bulge was explored, but with the average sizes of bulges ($\sim$1 kpc) and the resolution of GAMA, the S\'{e}rsic index would be poorly constrained (given the quality of sky subtraction and accuracy of the PSF). This poor constraint would add sufficient degeneracy to the fit, and hence for this work we have deemed it more favourable to simply fix the S\'{e}rsic index of the bulge in the BD mode. 
This is consistent with approaches taken in the literature  \citep[for example][]{simard2011, barsanti2021a}, although we note this is simpler than studies that fit a free S\'{e}rsic index for the bulge \citep[for example][]{dimauro2018, cook2019, costantin2021, casura2022, haussler2022}.

\subsubsection{SED-fitting models}

To constrain the relative brightness of the two-dimensional models across different wavelengths, an SED for each component is generated, in much the same way as traditional SED fitting. 

The SED modelling implementation used for all four \textsc{ProFuse} configurations is near identical, and follows the \textsc{ProSpect} modelling approach used by \citet{bellstedt2020b} and \citet{bellstedt2021}. 
In short, a parametric star formation history is implemented, using the \texttt{massfunc\_snorm\_trunc} parametrisation, i.e. a skewed Normal star formation history with a truncation implemented in the early Universe (forcing the SFR to be 0 at the start of cosmic time). 
Metallicity is allowed to evolve linearly (meaning that the metallicity growth is mapped directly to the growth in stellar mass of the component, achieved using the \texttt{Zfunc\_massmap\_lin} parametrisation), with the final gas-phase metallicity for each component modelled as a free parameter. 
In two-component configurations, the star formation and metallicity histories are therefore entirely independent for each of the components. 
Because this star formation and metallicity implementation is identical to \textsc{ProSpect} implementations conducted in previous works \citep[see][]{bellstedt2020b, bellstedt2021, thorne2022, thorne2023}, we refer the reader to those papers for a detailed presentation of the interplay between metallicity and age, noting that the accurate recovery of the cosmic star formation history by \citet{bellstedt2020b} using this metallicity evolution implementation demonstrated that the ages are not systematically biased due to this degeneracy. 

Because far-IR data are not being used in this \textsc{ProFuse} implementation, there is very little constraining power for the various dust parameters specified in \textsc{ProSpect}. For this reason, we fix the dust parameters to typical galaxy values as identified by \citet{thorne2022}. The exception is the FS configuration, where we allow the tau parameters (relating to dust opacity) to be free (possible due to the smaller number of free parameters). However even in this configuration the alpha parameters (which relate to the dust temperature) are fixed, as they are unconstrainable without far-IR data.

All of the relevant fixed and free parameters, as well as their values and fitting ranges, have been provided for the four model configuration in Tables \ref{tab:FSparams}-\ref{tab:DDparams}. The total number of free parameters is 13 for FS, 16 for BD, 15 for PD, and 18 for DD. 

\begin{table}
	\centering
	\caption[FS Parameters]{Parameters fitted for the FS configuration, with a total of 13 free parameters. }
	\label{tab:FSparams}
	\begin{tabular}{@{}l | cc |  c}
		\hline
		Parameter & Log & Range/Value &Units\\
		\hline
		\hline
		\multicolumn{4}{l}{\textit{SFH parameters}} \\[3pt]
		\hline
		mSFR  & Yes & [$-4$, 3] & $\rm M_{\odot}/{\rm yr}$ \\
		mpeak  & No & [$-2$, $13.4-t_{\rm lb}$]  & Gyr\\
		mperiod  & Yes & [$-1$, 1] & Gyr  \\
		mskew  & No & [$-1$, 1] & -- \\
		\hline
		\multicolumn{4}{l}{\textit{Metallicity parameters}} \\[3pt]
		\hline
		Zfinal  & Yes & [$-4$, $-1.3$]&  -- \\
		\hline
		\multicolumn{4}{l}{\textit{Dust parameters}} \\[3pt]
		\hline
		$\tau_{\rm birth}$  & Yes & [$-2.5$, 1]& -- \\
		$\tau_{\rm screen}$  & Yes & [$-2.5$, 1] & --\\
		$\alpha_{\rm birth}$  & -- & 1 &  --\\
		$\alpha_{\rm screen}$  & -- & 3  & --\\
		${\rm pow}_{\rm birth}$  & -- & -0.7 & --  \\
		${\rm pow}_{\rm screen}$  & --& -0.7 & --  \\
		\hline
		\multicolumn{4}{l}{\textit{Structural decomposition parameters}} \\[3pt]
		\hline
		x   & No &[x$_{\rm input}-10$, x$_{\rm input}+10$]& pix \\
		y   & No & [y$_{\rm input}-10$, y$_{\rm input}+10$]& pix \\
		$n$   & Yes & [$\log_{10}(0.5)$, $\log_{10}(8)$]& -- \\
		axrat   & Yes & [-2, 0]& -- \\
		Re   & Yes & [1, $\log_{10}$(image max)]& pix \\
		ang   & No & [-180, 360]& deg \\
		\hline
	\end{tabular}
\end{table}

\begin{table}
	\centering
	\caption[BD Parameters]{Parameters fitted for the BD configuration, with a total of 16 free parameters. Fixed tau parameters are based on median values from \citet{thorne2021}, and no dust is assumed present in the bulge.}  
	\label{tab:BDparams}
	\begin{tabular}{@{}l | cc |  c}
		\hline
		Parameter  & Log & Range/Value &Units\\
		\hline
		\hline
		\multicolumn{4}{l}{\textit{SFH parameters}} \\[3pt]
		\hline
		disk mSFR  & Yes & [$-4$, 3] & $\rm M_{\odot}/{\rm yr}$\\
		disk mpeak  & No & [$-2$, $13.4-t_{\rm lb}$]  & Gyr\\
		disk mperiod  & Yes & [$-1$, 1] & Gyr  \\
		disk mskew  & No & [$-1$, 1] & -- \\
		bulge mSFR  & Yes & [$-4$, 3] & $\rm M_{\odot}/{\rm yr}$\\
		bulge mpeak  & No & [$-2$, $13.4-t_{\rm lb}$]  & Gyr\\
		bulgek mperiod  & Yes & [$-1$, 1] & Gyr  \\
		bulge mskew  & No & [$-1$, 1] & -- \\
		\hline
		\multicolumn{4}{l}{\textit{Metallicity parameters}} \\[3pt]
		\hline
		disk Zfinal  & Yes & [$-4$, $-1.3$]&  -- \\
		bulge Zfinal  & Yes & [$-4$, $-1.3$]&  -- \\
		\hline
		\multicolumn{4}{l}{\textit{Dust parameters}} \\[3pt]
		\hline
		disk $\tau_{\rm birth}$  & -- & 0.63 & -- \\
		disk $\tau_{\rm screen}$  & -- & 0.16 & --\\
		disk $\alpha_{\rm birth}$  & -- & 1&  --\\
		disk $\alpha_{\rm screen}$ & -- & 3  & --\\
		disk ${\rm pow}_{\rm birth}$  & -- &-0.7& --  \\
		disk ${\rm pow}_{\rm screen}$  & --&-0.7 & --  \\
		bulge $\tau_{\rm birth}$  & -- &0.63& -- \\
		bulge $\tau_{\rm screen}$  & -- & 0 & --\\
		bulge $\alpha_{\rm birth}$ & -- & 1&  --\\
		bulge $\alpha_{\rm screen}$  & -- & 3  & --\\
		bulge ${\rm pow}_{\rm birth}$  & -- &-0.7& --  \\
		bulge ${\rm pow}_{\rm screen}$  & --&-0.7 & --  \\
		\hline
		\multicolumn{4}{l}{\textit{Structural decomposition parameters}} \\[3pt]
		\hline
		x   & No &[x$_{\rm input}-10$, x$_{\rm input}+10$]& pix \\
		y   & No & [y$_{\rm input}-10$, y$_{\rm input}+10$]& pix \\
		disk $n$   & -- & 1& -- \\
		disk Re   & Yes & [1, $\log_{10}$(image max)]& pix \\
		disk axrat   & Yes & [-2, 0]& -- \\
		disk ang   & No & [-180, 360]& deg \\
		bulge $n$   & -- & 4 & -- \\
		bulge Re   & Yes & [1, $\log_{10}$(image max)]& pix \\
		bulge axrat   & -- & 1 & -- \\
		bulge ang   & -- & 0 & deg \\
		\hline
	\end{tabular}
\end{table}

\begin{table}
	\centering
	\caption[PD Parameters]{Parameters fitted for the PD configuration, with a total of 15 free parameters. }
	\label{tab:PDparams}
	\begin{tabular}{@{}l | cc |  c}
		\hline
		Parameter & Log & Range/Value &Units\\
		\hline
		\hline
		\multicolumn{4}{l}{\textit{SFH parameters}} \\[3pt]
		\hline
		disk mSFR &  Yes & [$-4$, 3] & $\rm M_{\odot}/{\rm yr}$ \\
		disk mpeak &  No & [$-2$, $13.4-t_{\rm lb}$]  & Gyr\\
		disk mperiod  & Yes & [$-1$, 1] & Gyr  \\
		disk mskew & No & [$-1$, 1] & -- \\
		bulge mSFR  & Yes & [$-4$, 3] & $\rm M_{\odot}/{\rm yr}$ \\
		bulge mpeak  & No & [$-2$, $13.4-t_{\rm lb}$]  & Gyr\\
		bulge mperiod  & Yes & [$-1$, 1] & Gyr  \\
		bulge mskew  & No & [$-1$, 1] & -- \\
		\hline
		\multicolumn{4}{l}{\textit{Metallicity parameters}} \\[3pt]
		\hline
		disk Zfinal  & Yes & [$-4$, $-1.3$]&  -- \\
		bulge Zfinal  & Yes & [$-4$, $-1.3$]&  -- \\
		\hline
		\multicolumn{4}{l}{\textit{Dust parameters}} \\[3pt]
		\hline
		disk $\tau_{\rm birth}$  & -- & 0.63& -- \\
		disk $\tau_{\rm screen}$  & -- & 0.16 & --\\
		disk $\alpha_{\rm birth}$  & -- & 1 &  --\\
		disk $\alpha_{\rm screen}$  & -- & 3  & --\\
		disk ${\rm pow}_{\rm birth}$  & -- &-0.7& --  \\
		disk ${\rm pow}_{\rm screen}$  & --&-0.7 & --  \\
		bulge $\tau_{\rm birth}$  & -- & 0.63& -- \\
		bulge $\tau_{\rm screen}$  & -- & 0& --\\
		bulge $\alpha_{\rm birth}$  & -- & 1 &  --\\
		bulge $\alpha_{\rm screen}$  & -- & 3 & --\\
		bulge ${\rm pow}_{\rm birth}$  & -- &-0.7& --  \\
		bulge ${\rm pow}_{\rm screen}$  & --&-0.7 & --  \\
		\hline
		\multicolumn{4}{l}{\textit{Structural decomposition parameters}} \\[3pt]
		\hline
		x  &  No &[x$_{\rm input}-10$, x$_{\rm input}+10$]& pix \\
		y  &  No & [y$_{\rm input}-10$, y$_{\rm input}+10$]& pix \\
		disk $n$   & -- & 1 & -- \\
		disk Re   & Yes & [1, $\log_{10}$(image max)]& pix \\
		disk axrat   & Yes & [-2, 0]& -- \\
		disk ang   & No & [-180, 360]& deg \\
		\hline
	\end{tabular}
\end{table}

\begin{table}
	\centering
	\caption[DD Parameters]{Parameters fitted for the DD configuration, with a total of 18 free parameters. Note that we always refer to disk 1 as the disk with the smaller $Re$.} 
	\label{tab:DDparams}
	\begin{tabular}{@{}l | cc |  c}
		\hline
		Parameter & Log & Range/Value &Units\\
		\hline
		\hline
		\multicolumn{4}{l}{\textit{SFH parameters}} \\[3pt]
		\hline
		disk 1 mSFR & Yes & [$-4$, 3] & $\rm M_{\odot}/{\rm yr}$ \\
		disk 1 mpeak &  No & [$-2$, $13.4-t_{\rm lb}$]  & Gyr\\
		disk 1 mperiod &  Yes & [$-1$, 1] & Gyr  \\
		disk 1 mskew &  No & [$-1$, 1] & -- \\
		disk 2 mSFR &  Yes & [$-4$, 3] & $\rm M_{\odot}/{\rm yr}$ \\
		disk 2 mpeak &  No & [$-2$, $13.4-t_{\rm lb}$]  & Gyr\\
		disk 2 mperiod &  Yes & [$-1$, 1] & Gyr  \\
		disk 2 mskew &  No & [$-1$, 1] & -- \\
		\hline
		\multicolumn{4}{l}{\textit{Metallicity parameters}} \\[3pt]
		\hline
		disk Zfinal &  Yes & [$-4$, $-1.3$]&  -- \\
		bulge Zfinal &  Yes & [$-4$, $-1.3$]&  -- \\
		\hline
		\multicolumn{4}{l}{\textit{Dust parameters}} \\[3pt]
		\hline
		disk 1 $\tau_{\rm birth}$ &  -- & 0.63& -- \\
		disk 1 $\tau_{\rm screen}$ &  -- &0 & --\\
		disk 1 $\alpha_{\rm birth}$ &  -- & 1 &  --\\
		disk 1 $\alpha_{\rm screen}$ &  -- & 3  & --\\
		disk 1 ${\rm pow}_{\rm birth}$ &  -- &-0.7& --  \\
		disk 1 ${\rm pow}_{\rm screen}$ & --&-0.7 & --  \\
		disk 2 $\tau_{\rm birth}$ &  -- & 0.63 & -- \\
		disk 2 $\tau_{\rm screen}$ & -- & 0.16 & --\\
		disk 2 $\alpha_{\rm birth}$ &  -- & 1&  --\\
		disk 2 $\alpha_{\rm screen}$ & -- & 3 & --\\
		disk 2 ${\rm pow}_{\rm birth}$  & -- &-0.7& --  \\
		disk 2 ${\rm pow}_{\rm screen}$  & --&-0.7 & --  \\
		\hline
		\multicolumn{4}{l}{\textit{Structural decomposition parameters}} \\[3pt]
		\hline
		x  &  No &[x$_{\rm input}-10$, x$_{\rm input}+10$]& pix \\
		y  &  No & [y$_{\rm input}-10$, y$_{\rm input}+10$]& pix \\
		disk 1 $n$  &  -- & 1 & -- \\
		disk 1 Re  &  Yes & [1, $\log_{10}$(image max)]& pix \\
		disk 1 axrat  &  Yes & [-2, 0]& -- \\
		disk 1 ang  &  No & [-180, 360]& deg \\
		disk 2 $n$  &  -- & 1 & -- \\
		disk 2 Re  &  Yes & [1, $\log_{10}$(image max)]& pix \\
		disk 2 axrat  &  Yes & [-2, 0]& -- \\
		disk 2 ang  &  No & [-180, 360]& deg \\
		\hline
	\end{tabular}
\end{table}

\subsubsection{Dominant uncertainty}

\begin{figure}
	\centering
	\includegraphics[width=85mm]{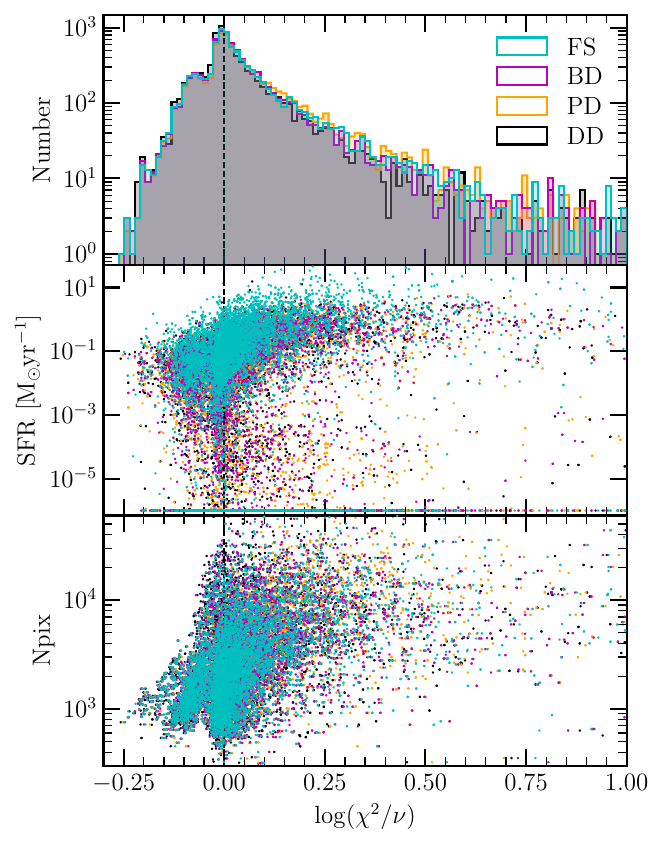}
	\caption{Top panel: Distribution of reduced chi-squared values for the four different model configurations implemented.
	Middle panel: Variation of reduced chi-squared values with the total modelled SFR (all galaxies with SFR $<10^{-6}\,\rm{M}_{\odot} \rm{yr}^{-1}$ have been plotted at $10^{-6}\,\rm{M}_{\odot} \rm{yr}^{-1}$ for visualisation purposes only). 
	Lower panel: Variation of reduced chi-squared values with the number of pixels in the object segment.  }  
	\label{fig:Chi2}
\end{figure}

With a fitting process as intricate as the one employed in this work, there are a number of potential sources of uncertainty. 
In general, the dominant uncertainty is the limitations of the model, as opposed to the quality of the data. 
This is shown in Fig. \ref{fig:Chi2}, where we show the distribution of reduced $\chi^2$ values of the sample for the different model configurations. 
The vast majority of galaxies have a $\chi^2/\nu$ value around 1, indicating a good fit of the model to the data. 
There is a subtle skewing of galaxies with values greater than 1 though, indicating in this regime that the uncertainty of the fits is dominated by model systematics. 
While there are galaxies with $\chi^2/\nu < 1$, these are the galaxies with low SFRs (shown in the middle panel) and lower numbers of pixels (lower panel).
We note that a visual inspection of the fits for these galaxies generally shows good fits with little residual features, an indication that the complexity of our models is likely overfitting the data here. 
Again, this shows that the dominant uncertainty comes from limitations in the model, rather than the data quality. 
We note that a simple cut based on source magnitude or pixel number does not a priori identify these overfitted galaxies.   

\subsubsection{Unfitted galaxies}

There are 9 galaxies (corresponding to 0.13\% of the sample) for which there was no successful \textsc{ProFuse} fit, and these were therefore omitted from all subsequent analysis. Galaxies that have images missing in any bands, or that do not have sufficient nearby stars from which to measure the PSF will fail the default \textsc{ProFuse} pipeline.  This leaves a sample of 6,655 galaxies studied in this work. 

\subsection{Structural nomenclature}
\label{sec:nomenclature}

It is essential to be explicit about the nomenclature adopted in our work, as the usage of structural terms varies across the literature. The three terms that we adpot are \textit{disks}, \textit{bulges}, and \textit{spheroids}. 

\textbf{Disk} terminology is consistent with the common use, describing any flattened, circular structure. Disks can either be one part of a two-component system, or galaxies can be purely disks. 
\textbf{Bulge} is used to describe the ellipsoidal structure at the centre of a two-component system. 
\textbf{Spheroid}, in our work, is used to describe the single-component structures that are ellipsoidal. Note that with this usage, bulges are \textit{not} deemed to be a subset of spheroids (which is how the term is often used in the literature). 
In this sense, disks, bulges, and spheroids are not treated as overlapping categories, and are instead viewed as ``eigenstructures" of galaxies. 
We note that our spheroids are a much broader class than the ``elliptical" galaxy class. 

\subsection{Model Selection}

Regardless of the manner in which a galaxy decomposition is conducted, an essential part of the process is determining which model configuration best describes a galaxy's two-dimensional distribution of light. 
This can be the most challenging part of structural decomposition, as in many cases a single-component model can describe the light distribution of a galaxy just as well as a bulge plus disk model. 
What should then determine the choice?
This choice has frequently been made using visual classifications or inspections \citep[such as in component modelling by][]{gadotti2009}. 
While a completely numerical quantifier is desired (to remove subjective visual decisions), this approach has remained elusive \citep[although it has been described at length in works like][]{hashemizadeh2021}. 

Because reliable and purely numerical discriminators remain controversial, visual classifications have remained important in much morphology/structure-based work. 
As datasets are increasing in size dramatically though, requiring visual inspections of galaxies is becoming increasingly unfeasible. 
An alternate approach increasingly being used is machine learning, where effectively a computer is trained to quickly and efficiently replicate a visual classification on a very large scale. 
Many galaxy morphological classifications have been made in this manner \citep[see for example][]{walmsley2022, li2023, fang2023, tian2023}. At this stage these classifications have not yet readily been used for much in the way of follow-on science, although an exception to this is \citet{cavanagh2023}, who used machine learning morphologies to study the evolution of lenticular galaxies with cosmic time. 
The benefit from a machine learning approach is that larger sample sizes can be processed. 
Machine learning morphologies will suffer the same potential biases that any visual classification scheme will, due do its inherently qualitative nature, and because nearly all machine learning classifictions have been trained on visually classified data. 

For our work, we apply a numeric quantifier from our \textsc{ProFuse} outputs to quantitatively categorise our sample into bulges, disks, and spheroids. 
Acknowledging however that any classification scheme (whether numerical or visual) still bears some ambiguity, a key aim of this work is to demonstrate the inherent uncertainties that still accompany any attempt to classify galaxies into structural classes.
We therefore also incorporate a number of different visual morphological classifications that have been conducted for this sample of galaxies. 
These different classification schemes are outlined in the following subsections. 

\subsubsection{\textsc{ProFuse} numerical model selection}
\label{sec:ModelSelection}

The numerical best model selection is expanded from \citet{robotham2022}. 
Using the parameters derived from each of the models, simple arguments are used to decide whether one model is clearly more physical (for example, if the B/T is very low, then there is motivation to assume that only a single component is required to model to the galaxy). 
In cases where clear physical motivation is not found, the Deviation Information Criterion (DIC) (which is smaller in the case of a better fit to the data) is used to determine preferred models. 

\begin{figure*}
	\centering
	\includegraphics[width=163mm]{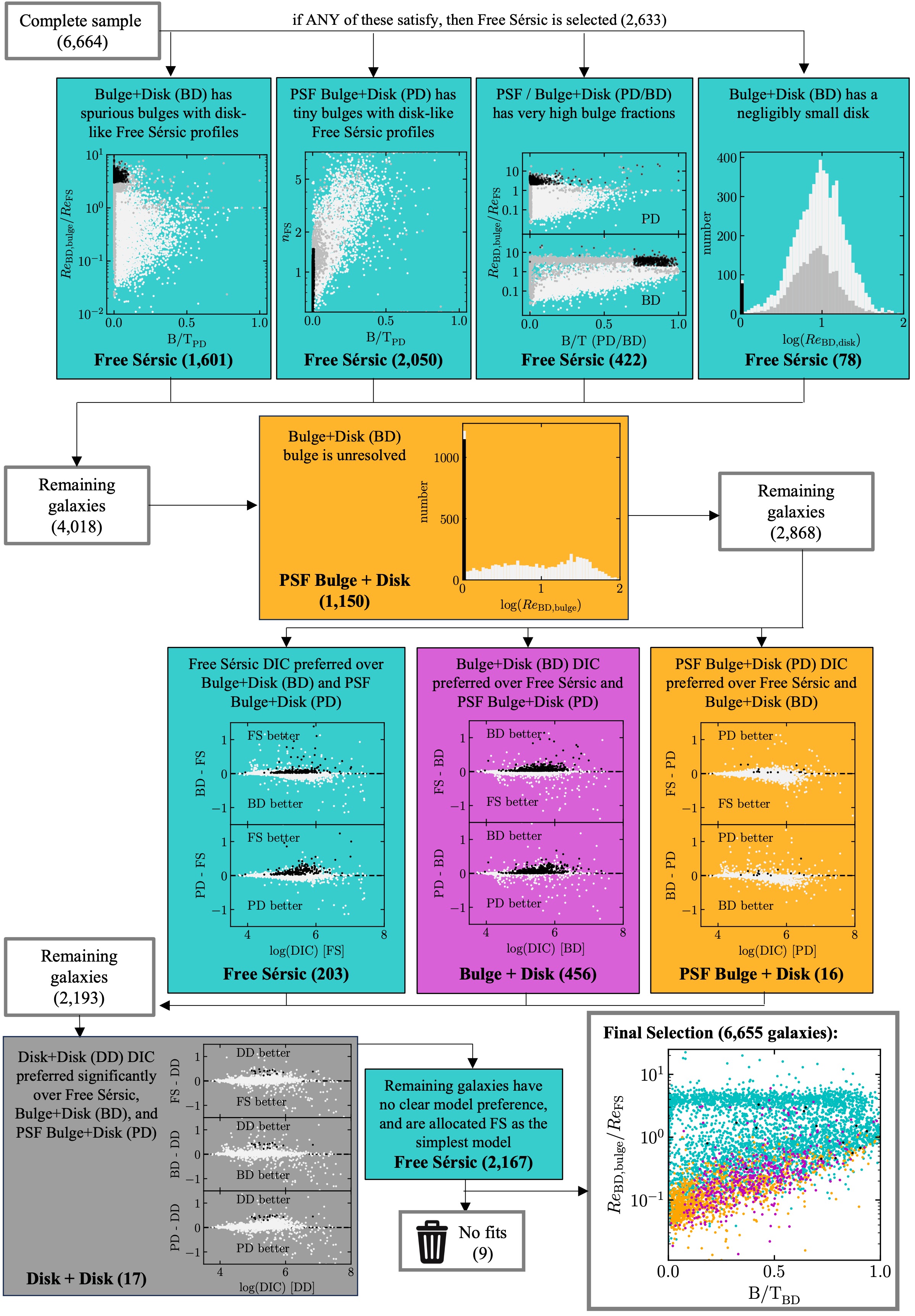}
	\caption{Visual representation of the numerical classification scheme. The number of sources classified in each step are indicated in each box, as is the relevant parameter space, with selected sources shown in black in each case.} 
	\label{fig:ClassificationFlowchart}
\end{figure*}

We represent this selection visually in Fig. \ref{fig:ClassificationFlowchart}. 

A galaxy is best-fitted by FS if it satisfies any of the following criteria:

1) if spurious bulges are fit using BD, for galaxies with disk-like profiles using FS
$$\left(\frac{Re_{\rm{BD}, \rm{bulge}}}{Re_{\rm{FS}}}>\textbf{3} \right) \&\left (\rm{B/T}_{\rm{PD}}<0.1\right) \& \left (n_{\rm{FS}}<1.5\right)$$ 

2) if negligibly tiny bulges are fit using PD, for galaxies with disk-like profiles using FS
$$\left(\rm{B/T}_{\rm{PD}}<0.01\right) \& \left(n_{\rm{FS}}<1.5\right) \& \left(n_{\rm{FS}}\geq 0.5\right)$$

3) if all two-component models show very high bulge fractions
$$\Bigl[ \left (\rm{B/T}_{\rm{PD}}>\textbf{0.7}\right) | \left(\rm{B/T}_{\rm{BD}}>\textbf{0.7}\right) \Bigr] \textbf{\&}  \left(\frac{Re_{\rm{BD}, \rm{bulge}}}{Re_{\rm{FS}}}>\textbf{3} \right)  $$

4) if the fitted disk in a BD system is negligibly small, indicating only the bulge component fits the whole galaxy
$$ Re_{\rm{BD}, \rm{disk}}=1 $$ 

Then, of the remaining unclassified galaxies, PD is selected if $ 0.1 <  Re_{\rm{BD}, \rm{bulge}} \leq 1.1 $ (measured in pixels), ensuring that a PSF bulge is only used if the fitted bulge in BD mode is significant, but smaller than the PSF. 

Finally, all remaining galaxies are classified based on the preferred DIC for each of the models (where a lower value indicates a better fit). 
The criteria for this are as follows. FS is selected if:
\begin{align*}
& \log(\rm{DIC}_{BD}/\rm{DIC}_{FS}) > \textbf{0.02} \,\,\,  \& \\
& \log(\rm{DIC}_{PD}/\rm{DIC}_{FS}) > \textbf{0.02} 
\end{align*}
BD is selected if:
\begin{align*}
& \log(\rm{DIC}_{FS}/\rm{DIC}_{BD}) > \textbf{0.02}\,\,\, \& \\
& \log(\rm{DIC}_{PD}/\rm{DIC}_{BD}) > 0.0
\end{align*}
PD is selected if:
\begin{align*}
& \log(\rm{DIC}_{FS}/\rm{DIC}_{PD}) > \textbf{0.03}\,\,\, \& \\
& \log(\rm{DIC}_{BD}/\rm{DIC}_{PD}) > 0.0
\end{align*}

Finally, of the remaining unclassified galaxies, DD is only selected for a small number of objects for which the model is significantly preferred:
\begin{align*}
& \log(\rm{DIC}_{FS}/\rm{DIC}_{DD}) > 0.2\,\,\,  \& \\
& \log(\rm{DIC}_{BD}/\rm{DIC}_{DD}) > 0.2\,\,\,  \& \\
& \log(\rm{DIC}_{PD}/\rm{DIC}_{DD}) > 0.2
\end{align*}

All remaining unclassified galaxies are selected to be FS, as the simplest model is assumed to be sufficient if no more complex models are preferred. 
The non-zero values in the above criteria are chosen to ensure that only models that are \textit{significantly} preferred according to their DIC are selected for more discrepant models. In the case of the FS selection, this ensures that potential BD sources are not missed (given that FS is selected as the final default regardless, if no other model is significantly preferred). PD and BD are so similar in terms of model complexity that this significance is not as important. The exact value was selected based on visual calibration (as was the case for each of the selection criteria). 

The above criteria have been slightly modified from the similar implementation by \citet{robotham2022}, to better catch two-component systems that were being flagged as single-component (based on extensive visual inspection and optimisation). The final DD criterion has been added (as the DD model had not yet been implemented for that work). The values that are different have been indicated using bold font. 

This approach results in a purely numerical mechanism of determining the structure of galaxies. 
In total, 5,016 galaxies were selected as best modelled by FS (free S\'{e}rsic), 456 by BD (bulge and disk), 1,166 by PD (PSF bulge and disk), and only 17 by DD (two disks). 


For all sources deemed best-described by a single component with a free S\'{e}rsic index (FS), it is necessary to make a separate decision on the type of structure that this component is consistent with, be they disk or spheroidal. 

For the sake of the structural analysis in this paper, we use the following criteria to classify these structures. 
If $n \leq 1.5$, the structure is a disk. 
If $n > 2.5$, the structure is a spheroid. 
If $1.5 < n \leq 2.5$, the structure is treated as ambiguous. 

Here, the ambiguous classification is simply included as an acknowledgement that the exact S\'{e}rsic cut used to separate disks and spheroids contributes to the classification uncertainty, and hence throughout the analysis in this work, the ambiguous sources will be used to contribute to the uncertainty measurements of each population.

\section{Comparing classification schemes}
\label{sec:ClassificationRegimeComparison}

The literature is littered with different methods of classifying galaxies into classes that describe their structure. 
To ensure that our results are not dependent on the choice of classification scheme, we include multiple visual classification schemes in the analysis. 
For the sake of making comparisons between the results derived from such classifications, it is necessary for us to comment on their relative agreement, or disagreement. 

\subsection{Visual morphological classifications}
\label{sec:VisualClassifications}

\def\SB{$\rm{Visual}_{\rm{SB}}$ }
\def\Hubble{$\rm{Visual}_{\rm{Hubble}}$ }
\def\DR4{$\rm{Visual}_{\rm{DR4}}$ }

A visual re-classification of the 6,664 galaxies of this work has been conducted (by SB) into structural classes that are equivalent to those defined by the automatic \textsc{ProFuse} classifications described in the previous section. 
The classes used in this classification are much broader and simpler than previously-used schemes, and are limited to three categories: disk, spheroid, and two-component galaxies. 
These have been called ``\SB classes" throughout the paper. 

These classes complement visual classifications that have been previously conducted on this sample by \citet{moffett2016a}  (``\Hubble Classes" hereafter), sorting the galaxies into their Hubble types. 
The Hubble classes used are ellipticals (\texttt{E}),  \texttt{S0-Sa},  \texttt{SB0-SBa},  \texttt{Sab-Scd},  \texttt{SBab-SBcd},  \texttt{Sd-Irr}, and Little Blue Spheroid (\texttt{LBS}) classes.  

As part of GAMA DR4, \citet{driver2022a} conducted an additional visual classification that was based on the visual structure of these galaxies, rather than the visual morphology. 
We refer to these classifications as the ``\DR4 Classes" hereafter. 
The DR4 classes separate galaxies into pure disks (\texttt{D}), ellipticals (\texttt{E}), compact sources (\texttt{C}), and then two-component systems, either with a compact bulge (\texttt{cBD}), or with a diffuse bulge (\texttt{dBD}). 
Galaxies that are too difficult to classify due to irregular structure such as from mergers are labelled ``hard" (\texttt{H}), or ``hard elliptical" (\texttt{HE}). 
These account for a tiny fraction of the sample, and as such are not significant in this analysis.

When presenting results from visual classification methods throughout this paper, we use properties derived from our \textsc{ProFuse} modelling. 
When visual classes have described a galaxy as having a single-component fit, we assign the FS properties to the galaxy. 
When a two-component system has instead been selected, then we can either assign the BD or PD properties to the galaxy. Wherever relevant for the duration of this paper, we present results with both options, to indicate the plausible uncertainty originating from model selection. We highlight that this is the most pessimistic uncertainty, as a likelihood criterion (as used in the automatic \textsc{ProFuse} classes) can be used to isolate the better-fitting description to shrink this uncertainty. 
This same approach is taken when presenting the bulge and disk properties using the \DR4 classes.

\subsection{Scheme Comparison}

\begin{figure*}
	\centering
	\begin{subfigure}[b]{1.0\textwidth}
	\includegraphics[width=180mm]{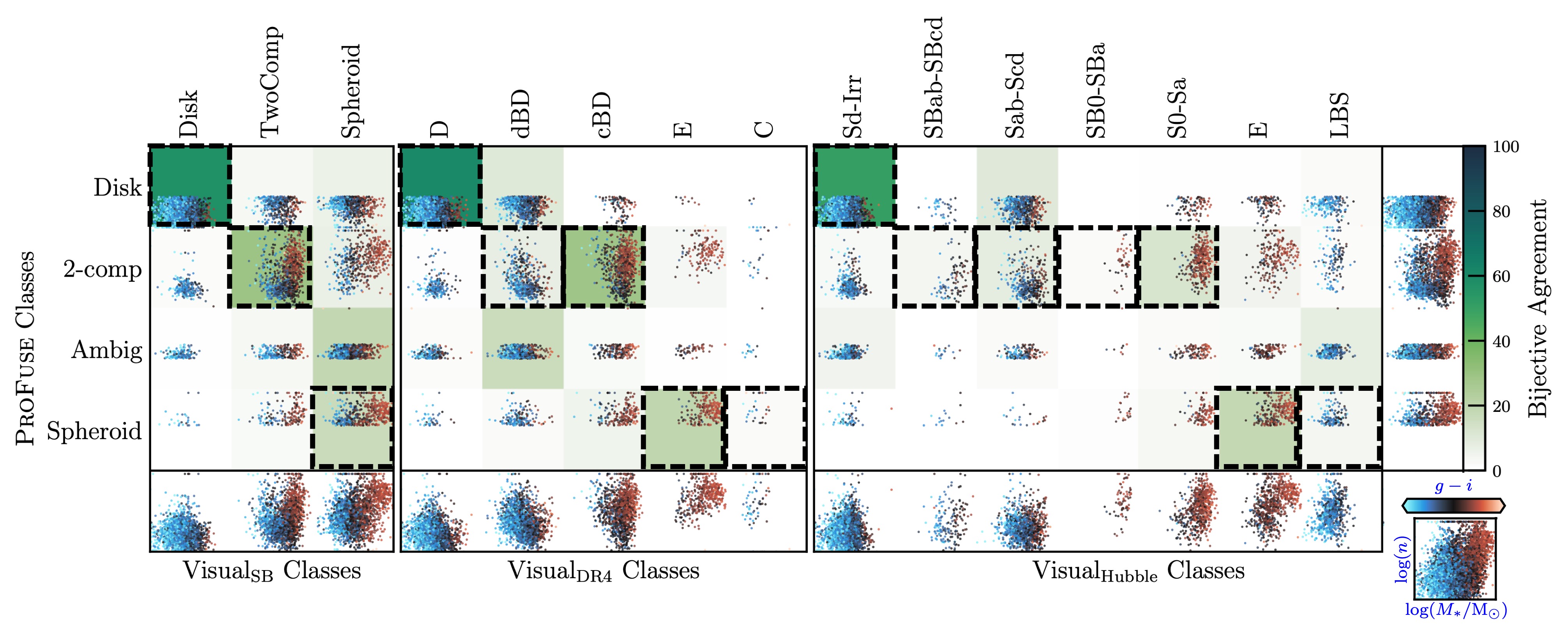}
	\end{subfigure}
	\begin{subfigure}[b]{1.0\textwidth}
	\includegraphics[width=180mm]{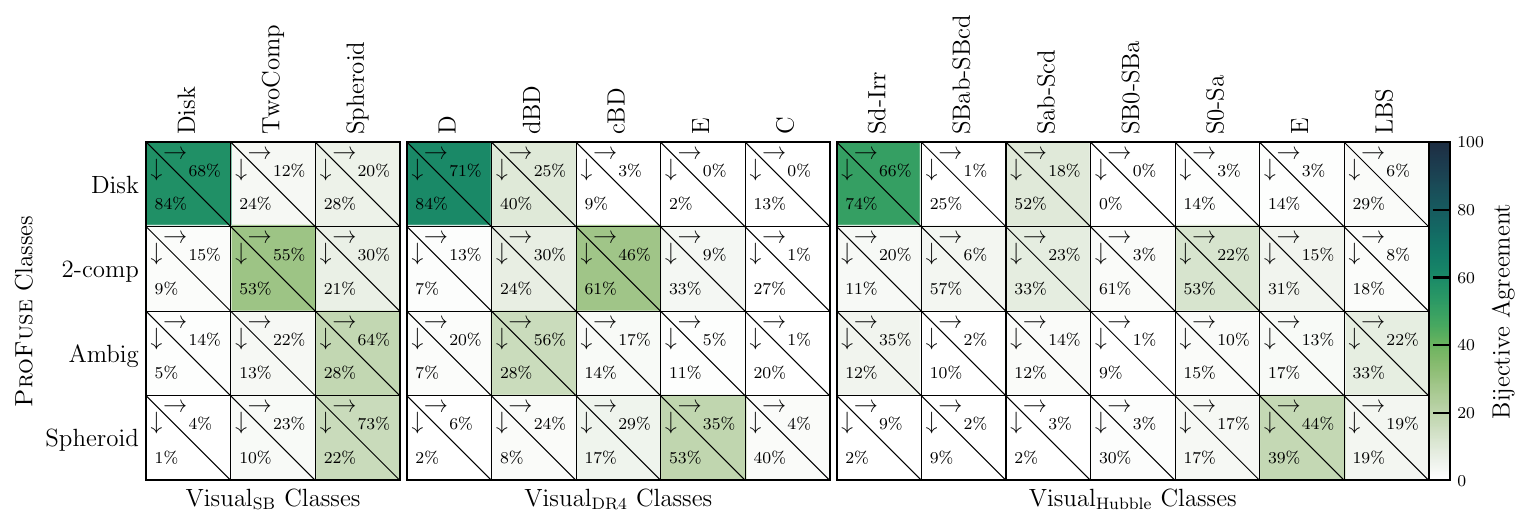}
	\end{subfigure}
	\caption{Top panel: Comparison of \textsc{ProFuse}-derived structural model, with the visual classifications from this work (left), the visual classification from \citet{driver2022a} (middle) and \citet{moffett2016a} (right).  
	For each subpopulation, the distribution of points in the mass--S\'{e}rsic index plane have been plotted, coloured by the $g - i$ colour, to give an idea of which galaxy properties are overwhelming each classification scheme. While not indicated on the plot for simplicity, the mass axis range is $ 7 < \log{M_*/\rm{M}_{\odot}} < 11.5$, the  S\'{e}rsic index axis range is $-0.3 < \log(n) < 0.95$, and the colour range is $0.3 < g-i < 1.5$.  
	The bijective agreement in each scheme is presented as the underlying green shading in each population (as indicated by the colourbar). 
	While the bijective agreement is overall similar between the scheme comparisons, it is interesting to note how the different conventions in visual classifications have had separate impacts in these comparisons. 
	Bottom panel: The fraction of sources overlapping, both according to vertical categories, and horizontal categories. For example, values with the horizontal arrow indicate the fraction of sources with that \textsc{ProFuse} class that overlap with each of the visual classes, in each classification scheme.}
	\label{fig:ModelSelectionMatrix}
\end{figure*}

We show a comparison between our \textsc{ProFuse} model classifications and the \SB Classes,  \DR4 Classes, and \Hubble Classes in Fig. \ref{fig:ModelSelectionMatrix}. 
The bijective agreement (which combines the fractional occupation of every class along both the row and the column) between the \textsc{ProFuse} classes and each of the visual classifications is shown in Fig. \ref{fig:ModelSelectionMatrix}. The darker the matrix element is coloured (as indicated by the colourbar), the better the agreement. 
For each matrix class combination,  we plot the $\log(M_*/\rm{M}_{\odot})$ -- $\log(n)$ distribution of the subset, coloured by the $g-i$ colour (the relevant subset of the left-hand panel). 
For each matrix entry, a dashed box indicates regions that we would reasonably expect to be populated if the classification schemes were working perfectly as intended. 
The fact that the majority of these regions are highly populated generally suggests that the classification schemes are working fairly consistently. 

The lower panel of Fig. \ref{fig:ModelSelectionMatrix} shows the equivalent matrix comparison, where the individual agreement percentages are shown in each category (showing greater detail than simply the coloured bijective agreement). 

These figures are designed to provide insight into the potential differences in the classification schemes, and include an enormous amount of information. Therefore, an exhaustive analysis of each intersection in these comparison matrices is not the aim of this section. Rather, the below subsections summarise some of the major conclusions apparent from this complex comparison. 


\subsubsection{Classifying disks}
The overlap between \SB disks, and single component disks from \textsc{ProFuse} is high, with a bijective agreement value of 57\%. 
The top left box in the lower panel of Fig. \ref{fig:ModelSelectionMatrix} shows that 84\% of all visually determined disks according to the \SB classes are flagged as disks according to the ProFuse model classifications, however of the ProFuse class disks, 68\% are visually classed as disks. Through comparison with the top panel, it can be seen that the ProFuse disks contain a larger portion of higher-mass, redder objects, that have been visually classed as either two-component or spheroid systems, despite all having low Sers\'{i}c indices.

The agreement between visual disks from DR4, and single component disks from \textsc{ProFuse} is also high, with a bijective agreement value of 59.6\%.  
Of the 29\% of \textsc{ProFuse} disks that are not classed as DR4 disks, the majority of these fall into the \texttt{dBD} category (disk with a diffuse bulge), which makes sense as such two-component systems with non-compact bulges are likely best modelled by a single component in our \textsc{ProFuse} setup (this uncertainty is shown by the fact that DR4 \text{dBD} galaxies are fairly evenly divided between \textsc{ProFuse} Disk, 2-comp, and Ambig categories). 

The \Hubble class that is most similar to pure-disk systems is \texttt{Sd} (where there is no prominent bulge), and so it is reassuring to see that the bijective agreement between \textsc{ProFuse} Disk and \texttt{Sd-Irr} galaxies is also high (with a value of 50.3\%). 
It is notable that there is also a substantial overlap between pure disks and \texttt{Sab-Scd} systems (which are deemed to have smaller bulges), albeit smaller (9.6\%).

\subsubsection{Classifying spheroids and ellipticals}
There is overlap between \textsc{ProFuse} spheroids and \SB spheroids, DR4 ellipticals and \Hubble ellipticals (with bijective agreement values of 16.1\%, 19.1\% and 17.7\% respectively), which is to be expected. 
The fact that the bijective agreements are so much lower than those of the pure disk populations, suggest that classifying non-disk single-component systems is much more prone to classification uncertainty.
Both our \textsc{ProFuse} and \SB spheroidal populations span the full range of both stellar mass and colour, which is contrasted to both visual elliptical classes that tend to only include high-mass, red galaxies. 
This is an indication thst colour what a much more influential factor in the classification of galaxies according to the \DR4 and \Hubble schemes, unlike the structural-only schemes presented in this paper. 

We find that when comparing \textsc{ProFuse} and \SB spheroids with the \Hubble classes, the majority of these low-mass, blue spheroids are either labelled as \texttt{LBS} (little blue spheroid) galaxies, or a small number of \texttt{Sd-Irr} galaxies. 
When compared with the \DR4 classes, the low-mass blue spheroids are instead classified as \texttt{dBD} sources, which are regarded as two-component, but without a typical, compact bulge. 
This shows that our definition of spheroidal is a much broader class than simply the elliptical class. For this reason it is important not to use the ``Elliptical" label for our spheroid class. 

\subsubsection{Classifying two-component systems}
\textsc{ProFuse}-classified two-component systems tend to mostly overlap with either  \texttt{cBD} or  \texttt{dBD} galaxies in the \DR4 scheme (as should be expected). 
There is however a smaller fraction of \textsc{ProFuse}-classified two-component systems (23\%) that are visually classified as one of the single-component classes (\texttt{D}, \texttt{E}, or \texttt{C}), which is an indication of the automatic versus visual classification error. 

In the \Hubble scheme there is a slight preference for them to overlap with the \texttt{S0-Sa} class (which would make sense given these are disk galaxies with a prominent bulge), however the number of \Hubble classes that are populated by the \textsc{ProFuse}-classified two-component systems is large. 
As indicated by the dashed boxes, there are four \Hubble classes that would be reasonable to expect are two-component systems, namely  \texttt{SBab-SBcd},  \texttt{Sab-Scd},  \texttt{SB0-SBa}, and  \texttt{S0-Sa}, however 43\% of \textsc{ProFuse}-classified two-component systems are classified as either \texttt{Sd-Irr}, \texttt{E}, or \texttt{LBS}, showing that substantial uncertainty remains in the decision to label these sources as either a single- or  two-component system (greater than the disagreement between the \textsc{ProFuse} and \DR4 classifications). 

\subsubsection{Disagreement in classifications across schemes}
\begin{figure}
	\centering
	\includegraphics[width=85mm]{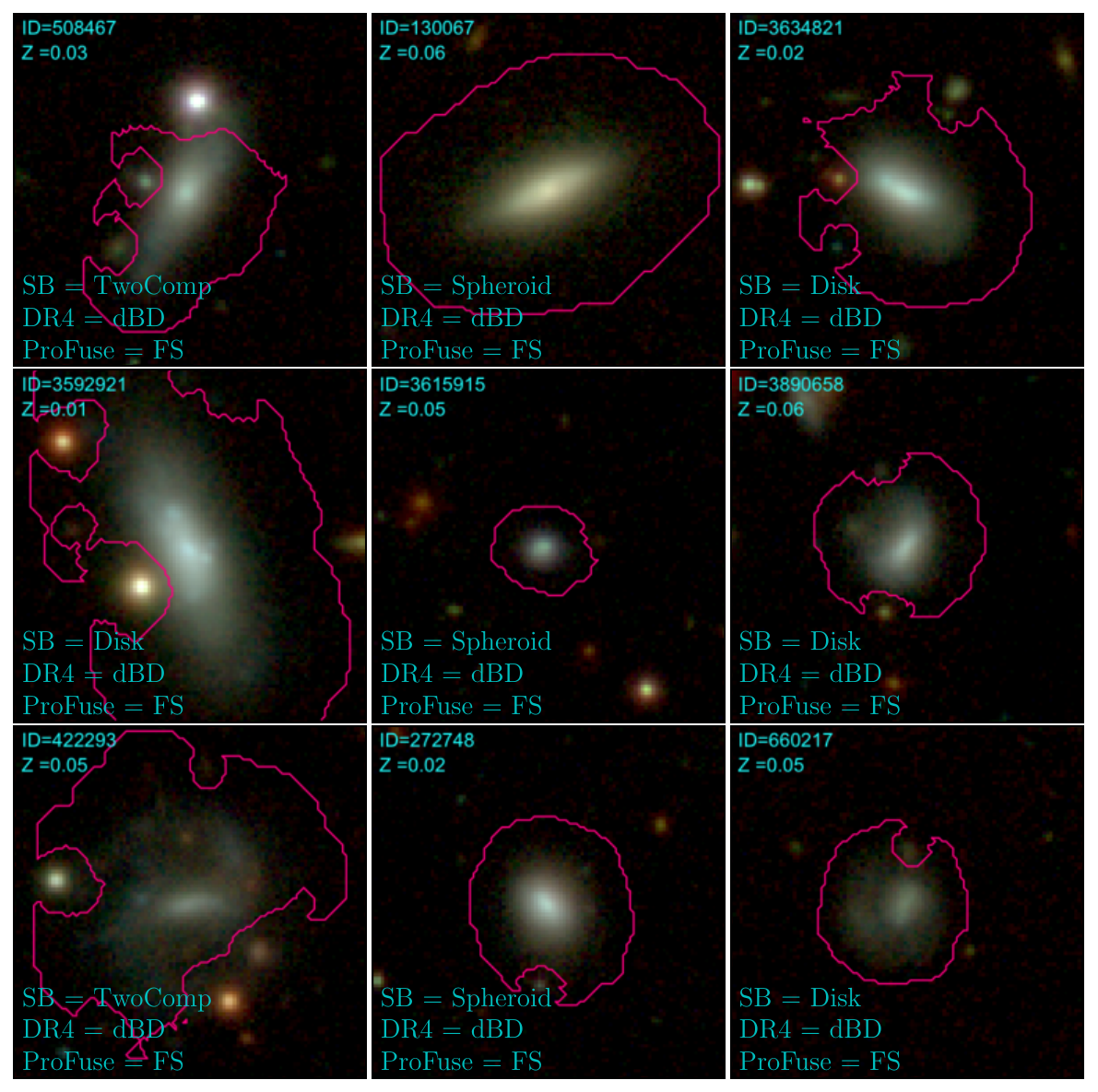}
	\caption{A random sample of objects that were classified as dBD by \citet{driver2022a}, but numerically classified by \textsc{ProFuse} as ambiguous, single-component systems. }
	\label{fig:DriverdBDProFuseAmbig}
\end{figure}

Regions that are more darkly shaded, but that do not have a dashed outline, represent regions where the classifications disagree. 
The vast majority of such regions actually fall within the \textsc{ProFuse}  \texttt{ambig} category, which accurately reflects the ambiguous nature of these sources. In the visual classifications of this work, they most closely relate to the visual spheroidal sources, suggesting that the $n$ cut used to separate disks and spheroids in the \textsc{ProFuse} classifiers is perhaps skewed slightly further toward disk than suggested by eye. 
These  \texttt{ambig} sources are most closely linked to \texttt{dBD} sources in the DR4 scheme. This suggests that the \textsc{ProFuse} classifications either miss diffuse bulges, or the visual classification is unnecessarily associating a bulge to a single-component system. 
Because bulges in our current implementation of \textsc{ProFuse} are always modelled as circular (in both the BD and PD models), it is possible that we are missing pseudobulge-style structures that are inherently elongated (as the two-component model may not be preferred in this scenario). 
A random sample of galaxies in this category is shown in Fig. \ref{fig:DriverdBDProFuseAmbig}.

\begin{figure}
	\centering
	\includegraphics[width=85mm]{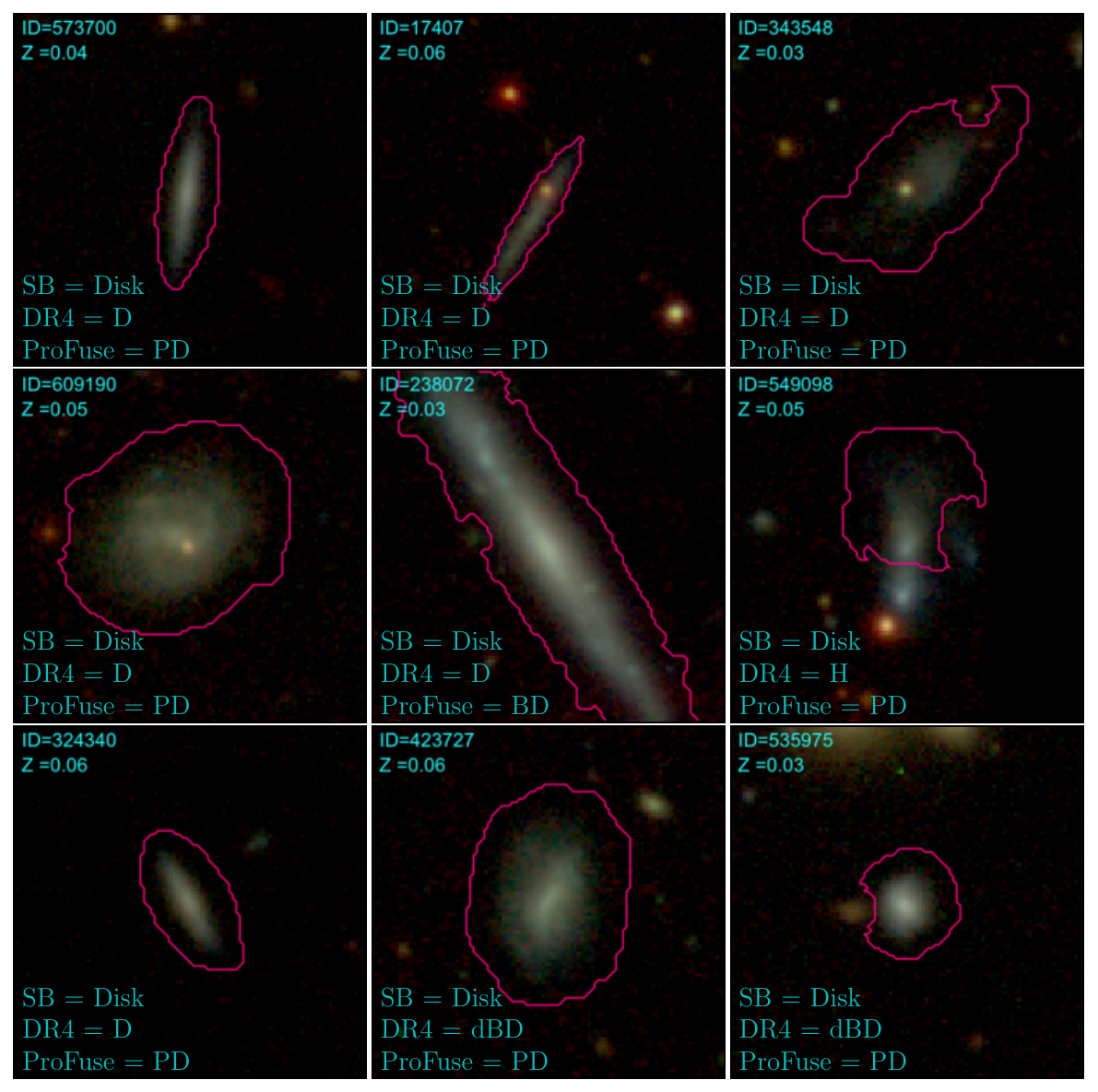}
	\caption{A random sample of objects that were visually classified as disks, but numerically classified by \textsc{ProFuse} as two-component systems. }
	\label{fig:SabineDiskProFuse2Comp}
\end{figure}

\begin{figure}
	\centering
	\includegraphics[width=85mm]{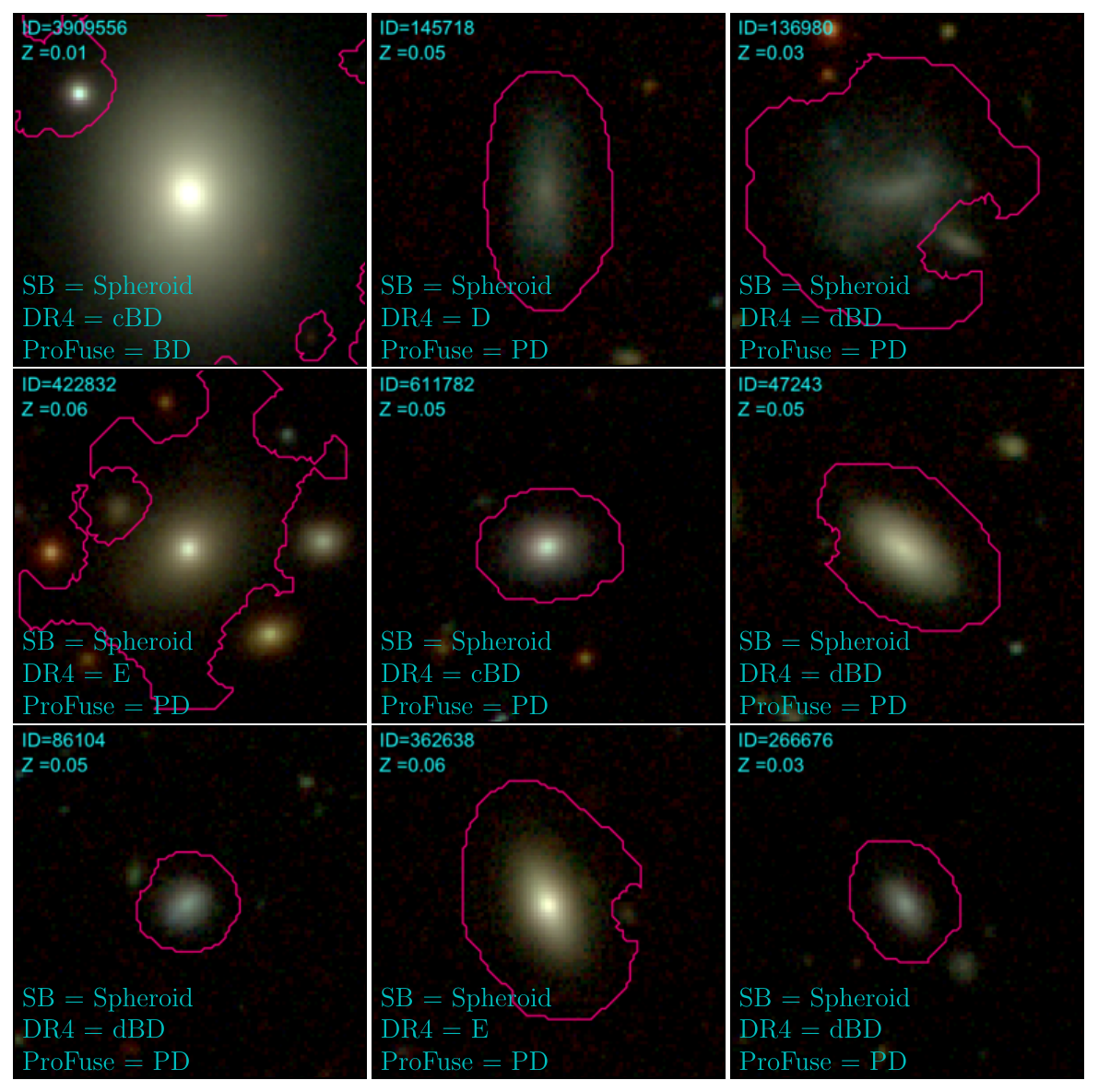}
	\caption{A random sample of objects that were visually classified as spheroids, but numerically classified by \textsc{ProFuse} as two-component systems.  }
	\label{fig:SabineSpheroidProFuse2Comp}
\end{figure}

Other categories that are perhaps surprisingly populated are the intersection between \textsc{ProFuse} two-component systems and each of the visual disks and spheroids. Random samples of each category  are presented in Figs. \ref{fig:SabineDiskProFuse2Comp} and \ref{fig:SabineSpheroidProFuse2Comp} respectively. A second component in each image is only subtle, however it is distinctly arguably that they do exist, suggesting that classification fatigue has resulted in these being missed when classifications were conducted by eye. In this scenario, the automatic classification from \textsc{ProFuse} seems to be superior. 

\subsubsection{Classification approach for subsequent analysis}
Our conclusion based on this comparison is that no single classification scheme appears to be objectively superior to the others, and there are benefits and drawbacks to each scheme.
For this reason, this work does not aim to provide a new classification that is superior to previous, but rather to use differing (but complementary) classification schemes to bracket the range of classifications, to estimate classification uncertainty. 
For the remainder of this paper, we will present results using each of the \textsc{ProFuse}, SB and DR4 structural classifications.

Throughout this paper, when we present statistics relating to the visual classes, we show outputs from our ProFuse modelling. For all single-component visual classes (\texttt{D},  \texttt{C},  \texttt{E}, and  \texttt{H}/ \texttt{HE}\footnote{Note that this category was not included in Fig. \ref{fig:ModelSelectionMatrix} simply because they are infrequently selected classes at our redshift range. }) we use the FS ProFuse model, and for all two-component classses (\texttt{cBD} and  \texttt{dBD}), we use the typical BD model. It is important to note therefore that the ``total" sample properties according to the two classes will therefore vary (even though the sample itself is the same).

\section{Results}
\label{sec:Results}

\subsection{Galaxy Fits}
\label{sec:GalaxyFits}

\begin{figure}
	\centering
	\includegraphics[width=85mm]{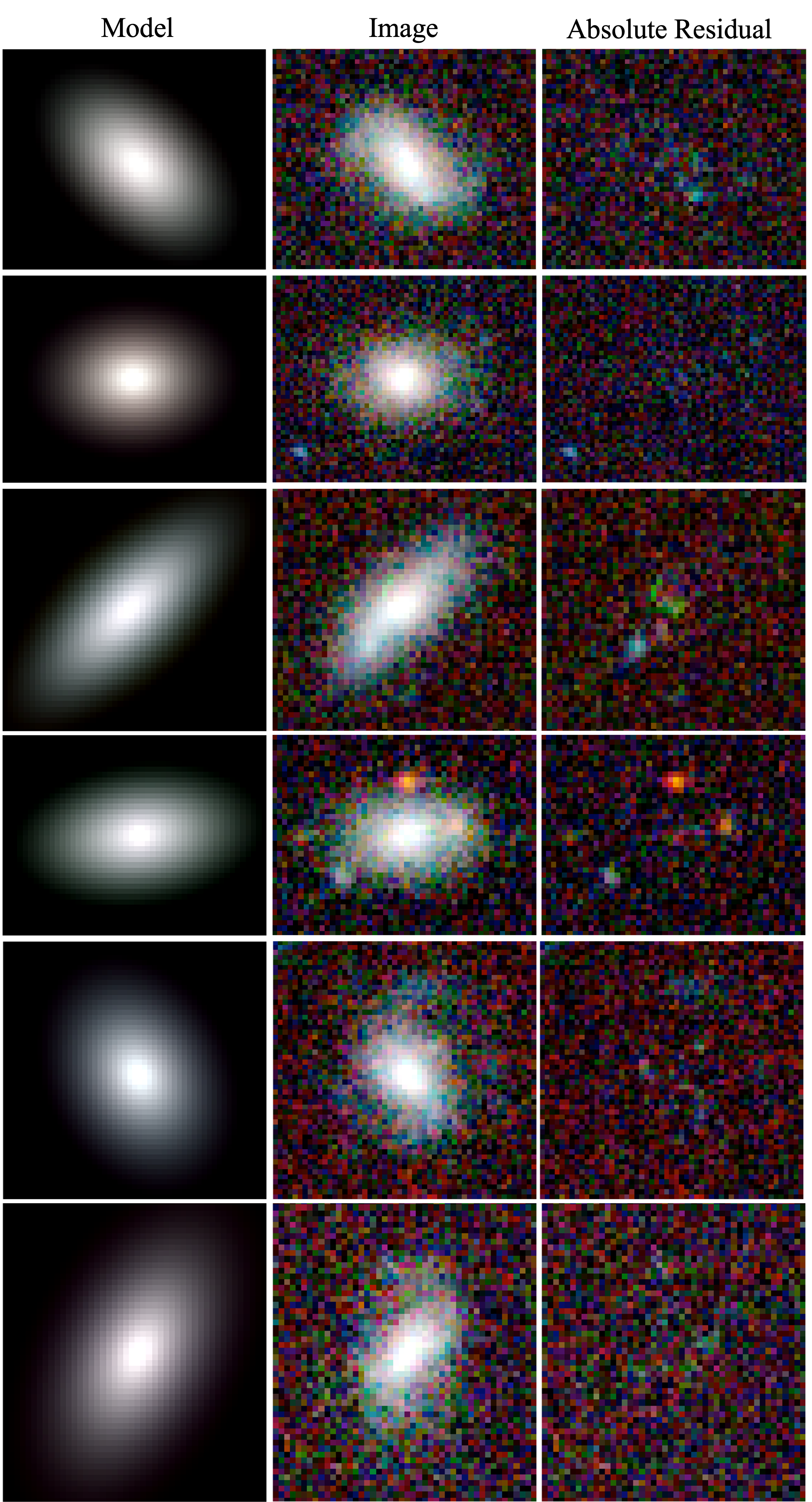}
	\caption{Compilation of RGB \textsc{ProFuse} outputs that show galaxies with no features in the residual, where the model provides a good description for the bulk of the light. The CATAIDs of presented galaxies are 15497, 15538, 17176, 17259, 22868, and 62785.  }
	\label{fig:RGBcompilation_noResidual}
\end{figure}

\begin{figure}
	\centering
	\includegraphics[width=85mm]{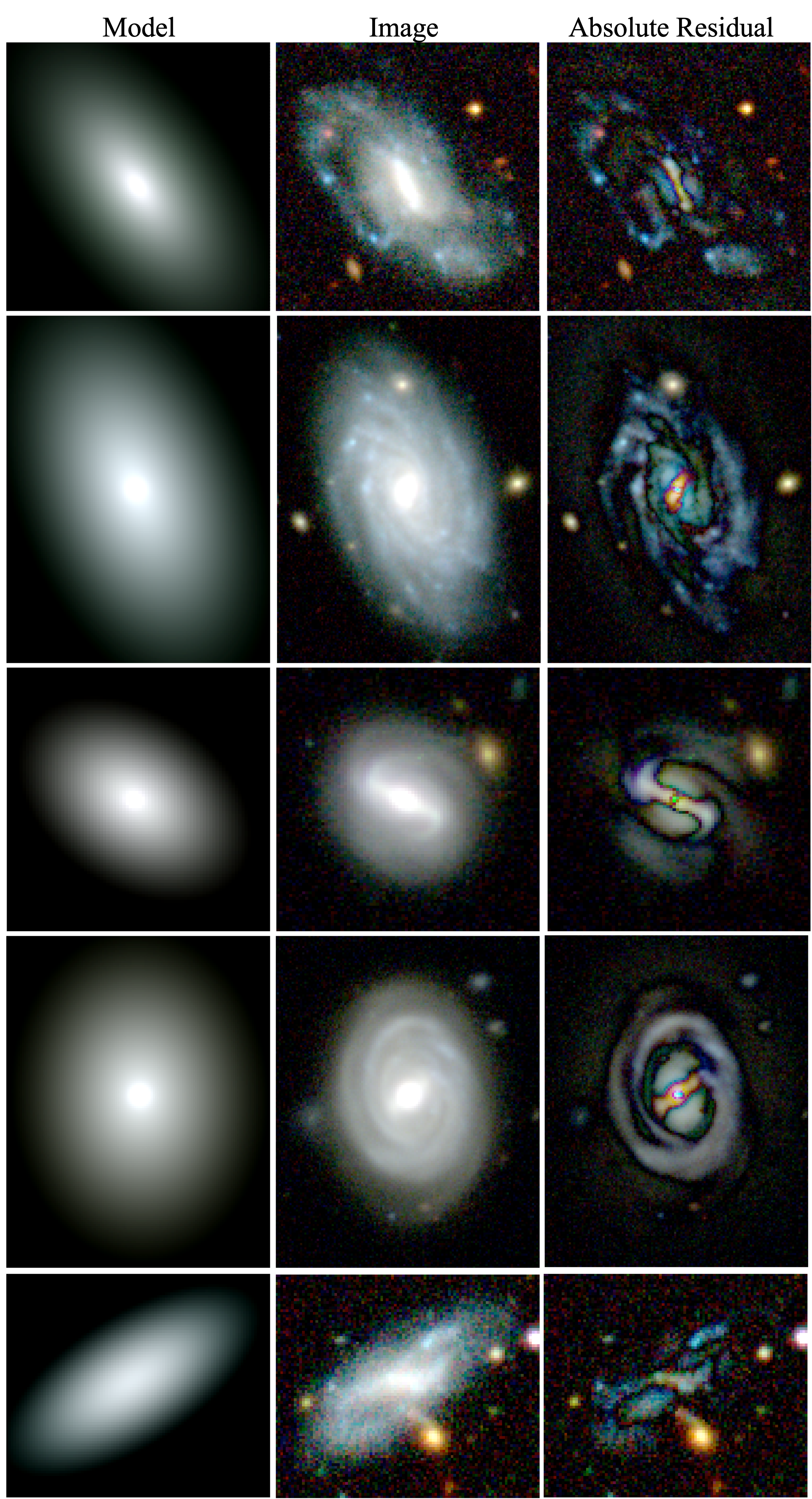}
	\caption{Compilation of RGB ProFuse ouptuts that have clear bar features in the residual. The CATAIDs of presented galaxies are 15218, 17314, 70929,  65406, and 23669.  }
	\label{fig:RGBcompilation_bars}
\end{figure}

An example of full \textsc{ProFuse} outputs can be seen in figure 8 of \citet{robotham2022}, which demonstrates how the SED and 2D fitting elements of this technique come together to produce an accurate model with a wealth of data for an individual galaxy and its structural components. 

As a brief demonstration of the galaxy fits achieved our \textsc{ProFuse} implementation in this work, we show three-band ($g$/$r$/$Z$) thumbnails of the best-fitting model and corresponding residual for a handful of sources in Figs. \ref{fig:RGBcompilation_noResidual}-\ref{fig:RGBcompilation_bars} (with a further selection presented in Appendix \ref{appendix:FittingOutputs}). Residuals shown are the absolute residual, highlighting both the negative and positive features. For each galaxy, the image, model and residual thumbnails are identically scaled, to make all visible features directly comparable. 

Fig. \ref{fig:RGBcompilation_noResidual} shows a handful of examples for which the \textsc{ProFuse} model results in little residual flux, largely due to the lack of visual substructures within the galaxies (seen in the right-most column of the figure). Because these galaxies are smaller on the sky, the lower effective resolution likely smooths over any substructure in the galaxies, making them easier to model. 

Structure types that have not been explicitly modelled in this work include components such as bars and spiral arms. Galaxies that have very strong bar features are shown in Fig. \ref{fig:RGBcompilation_bars}. Here, it is clear in the right column that there is a notable portion of the galaxy light that has not been well modelled by \textsc{ProFuse}. 
In particular, the bar is clearly visible in the residual images, and the red colour is dramatically different from the blue spiral arms, hinting at very distinct stellar populations within these structures. 
While we find that the total SED of the galaxy is still well approximated on average by \textsc{ProFuse} (and hence the global properties are not likely to be dramatically affected), the 2D model is non-optimal.
It has been shown by \citet{erwin2021} that a classical bulge plus disk decomposition of a barred galaxy can overestimate the spheroid components by factor of 4-100. 
Using the SDSS \citep[Sloan Digital Sky Survey, ][]{stoughton2002}, \citet{nair2010} estimate that 26\% of galaxies with $g<16$ have bars, which is higher than the results that come from a visual inspection of our colour residuals, in which we identify bars in 9\% of our sample in the same magnitude range. 
The resulting impact of neglecting bars in our fitting therefore has the potential to affect a significant fraction of our sample, although the light fraction of a bar in a galaxy can vary dramatically, so the true impact is difficult to estimate. 
Interestingly, while the residual structure for the galaxies in Fig. \ref{fig:RGBcompilation_bars} seems to be similar, a mix of models (BD, FS and PD) were selected for the galaxies themselves. 
This highlights that the presence of a bar does not tend to influence the automatic classifications from \textsc{ProFuse}, as bars are badly modelled by all model configurations in general. 

The residual features have a very strong colour signature. This indicates that an SED analysis of the residual features with \textsc{ProSpect} will already be possible with existing \textsc{ProFuse} outputs, allowing the stellar populations of features such as bars and spiral arms to be studied. 
For regular structures such as bars, it may be possible to include this as extra modelled structures in \textsc{ProFuse} itself, although care would need to be taken that there is sufficient quality in the imaging to constrain these extra free parameters. 
While it is beyond the scope of this work to conduct such an analysis, we feel it is a natural future extension. 
Additional samples of galaxies are presented in the same manner in Appendix \ref{appendix:FittingOutputs}, to demonstrate some of the interesting galaxy phenomena that are identifiable from this modelling method.

\subsection{Stellar mass distributions}

\begin{figure}
	\centering
	\includegraphics[width=85mm]{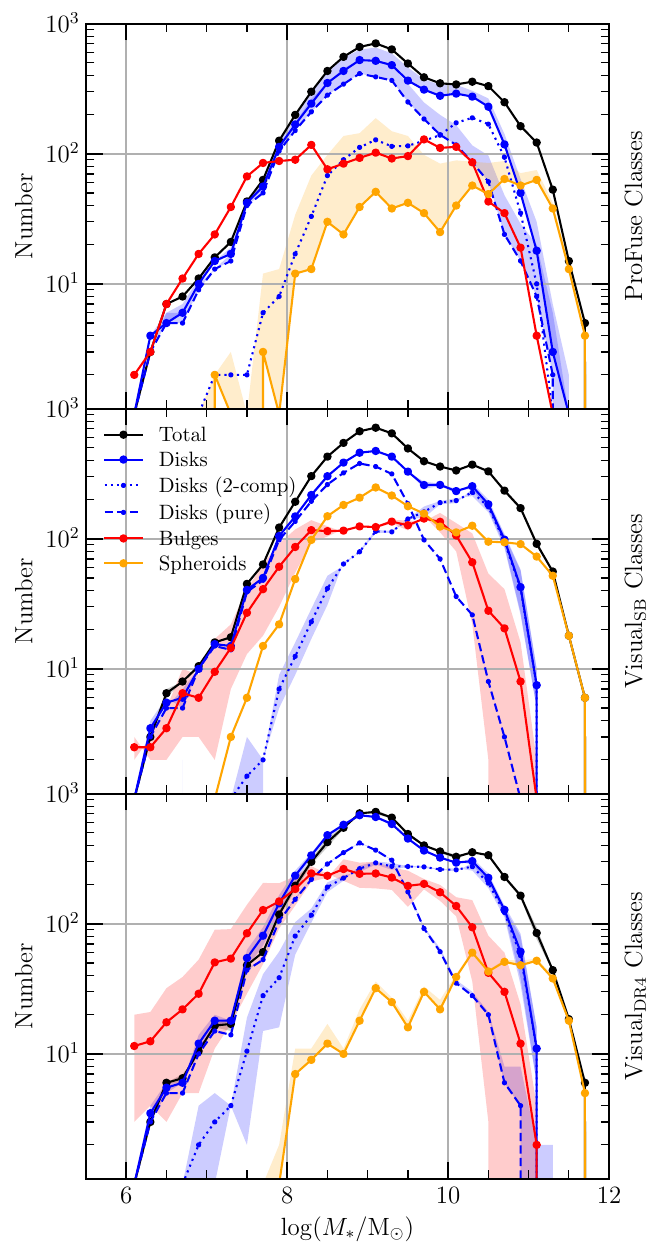}
	\caption{Stellar Mass distribution of all best-selected structural components in the top panel, \SB classes in the middle panel, and \DR4 classes in the bottom. The errorbars for the disk and spheroid populations in the top panel convey the range produced by the ambiguous population, reflecting the uncertainty introduced into these classifications based on the defined S\'{e}rsic index cut.  Errorbars in the middle and bottom panels correspond to the range in masses when two-component systems are either modelled by the BD or PD configurations. Finally in the bottom panel we also indicate the uncertainty from populations including the ``hard" class. }
	\label{fig:MassFunction_All}
\end{figure}

The stellar mass distributions of the full sample, as well as the disk, bulge, and spheroid populations are shown in Fig. \ref{fig:MassFunction_All}, for each of the three classification schemes used in this work. 
While the sample itself is identical in each of the schemes, the overall stellar mass distribution displays minor variation, thanks to the different combinations of ProFuse models that were selected to best describe each galaxy in each scheme. 
As mentioned in Sec. \ref{sec:VisualClassifications}, either PD or BD configurations could be assigned to galaxies that were visually deemed to be two-component systems. 
Rather than attempt assign a better model numerically to each individual galaxy, we conduct the analysis with both, which gives the most pessimistic estimate of the error. 
The uncertainty ranges in the bulge and two-component disk components in the middle and bottom panels of Fig. \ref{fig:MassFunction_All} are the extremes bounded by the scenarios where either all two-component systems are described by PD, or all two-component systems are described by BD. 
The use of PD generally results in a lower-mass bulge (because of the limited size that a bulge can have), and conversely the use of BD can produce higher-mass bulges. This is what is responsible for the substantial uncertainty in the bulge mass distribution at the low- and high-mass extremes. 
From the automatic \textsc{ProFuse} classes, it seems that lower-mass bulges are generally best described by the PD configuration, whereas high-mass bulges are best described by BD. 

There are significant differences that demonstrate the fundamentally different approaches used to classify galaxies. 
The double-peaked distribution of disks separates relatively cleanly between the pure disk and the two-component disk systems. This is true for all three classification schemes, although to different degrees. 
The \DR4 classes produce the highest low-mass peak in two-component disks, whereas the \SB classes have the lowest number of low-mass two-component disks (with the \textsc{ProFuse} classes somewhere in the middle). 
This behaviour is mirrored in the bulge population, with the \DR4 classes having the most low-mass bulges, the \SB classes the fewest, with the \textsc{ProFuse} classes somewhere in the middle. 
In general, it seems that the \textsc{ProFuse} classes produce a result that is the compromise between the different visual schemes at play in the \DR4 and \SB classes.

\begin{figure}
	\centering
	\includegraphics[width=85mm]{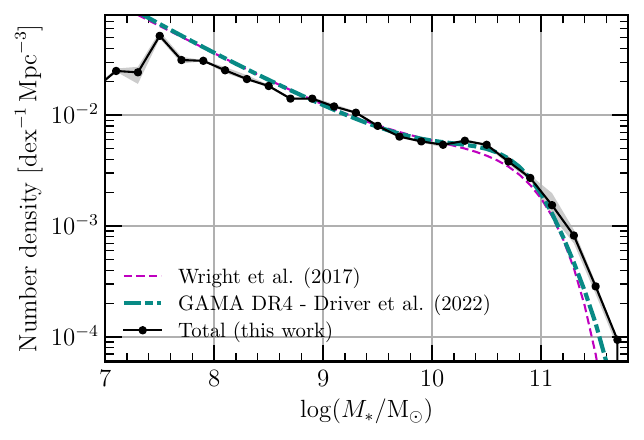}
	\caption{Total stellar mass function with the \textsc{ProFuse}-derived stellar masses. }
	\label{fig:MassFunction_total}
\end{figure}

To compare the overall mass distribution of our sample with previous work, we present the total mass function in Fig. \ref{fig:MassFunction_total}. This is compared against the recent mass function from \citet{driver2022}, and the previous GAMA mass function from \citet{wright2017}. 
We see that our mass function appears to be shifted slightly towards higher stellar masses than that of both  \citet{wright2017} and \citet{driver2022}. This is to be expected, given that the \textsc{ProFuse}-derived stellar masses are on average 0.1 dex larger than that of \textsc{ProSpect} (which is where the stellar masses originated from for the \citealt{driver2022} analysis). This offset is shown in fig. 11 of \citet{robotham2022}, with more detailed discussion than we repeat here.

\begin{figure}
	\centering
	\includegraphics[width=85mm]{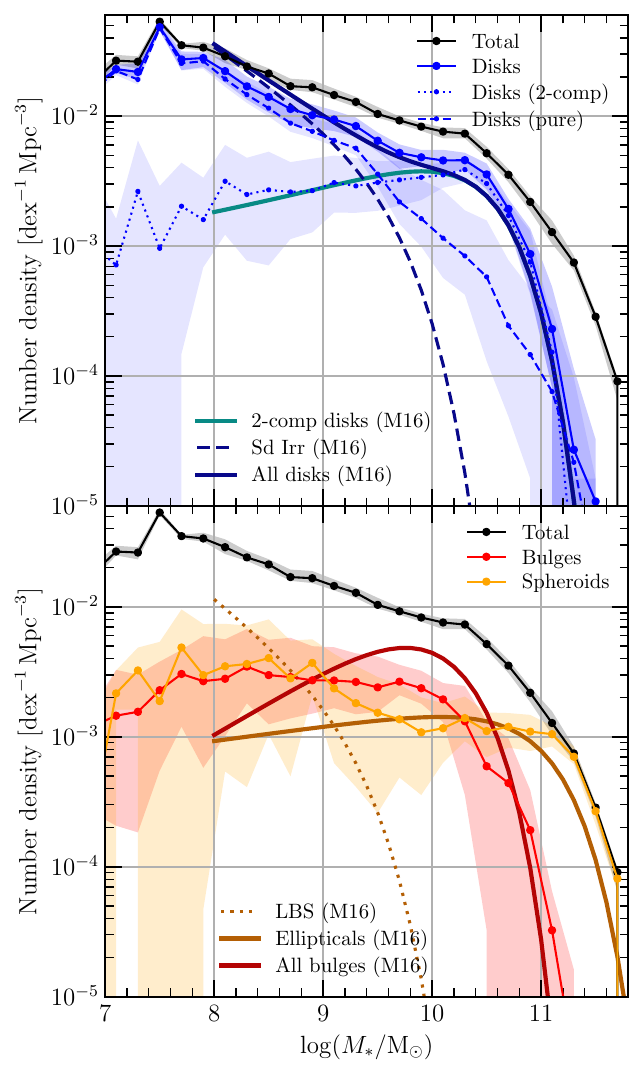}
	\caption{Per-component mass functions for each structure type, as compared with the relevant structural mass functions of \citet{moffett2016b}. Our points refer to the mass functions averages over all classification schemes, with the uncertainties indicating the plausible range covered by different galaxy structure classification methods. The top panel presents the disk-like components, whereas the bottom panel presents the bulge- and spheroid-like components. }
	\label{fig:MassFunction_combined}
\end{figure}

The mass function we derive when dividing by structure types is presented in Fig. \ref{fig:MassFunction_combined}, where the indicated uncertainty ranges convey the error produced through different classification approaches (compressing the information conveyed in the three panels of Fig. \ref{fig:MassFunction_All}).
Here, a completeness correction was conducted on a per-galaxy basis using the 1/$V_{\rm{max}}$ values for each galaxy computed using \textsc{ProSpect}. 
For ease of interpretation, the top panel of Fig. \ref{fig:MassFunction_combined} focuses on disk components, whereas the bottom panel focuses on the bulge and spheroid components. 
We include the Schechter fits to equivalent structure mass functions from \citet{moffett2016b} in coloured lines. 
The total disk mass functions are generally consistent, as is the two-component disk distribution. Equivalently, the spheroid distribution is also entirely consistent. 
Some differences exist when comparing the pure disk systems, where we note that our \textsc{ProFuse} pure disk mass function extends to much higher stellar masses than that of \citet{moffett2016b}. 
Similarly, our mass function for bulges is much less peaked, and contains a higher density of low-mass bulges than noted by \citet{moffett2016b}.

\begin{figure*}
	\centering
	\includegraphics[width=180mm]{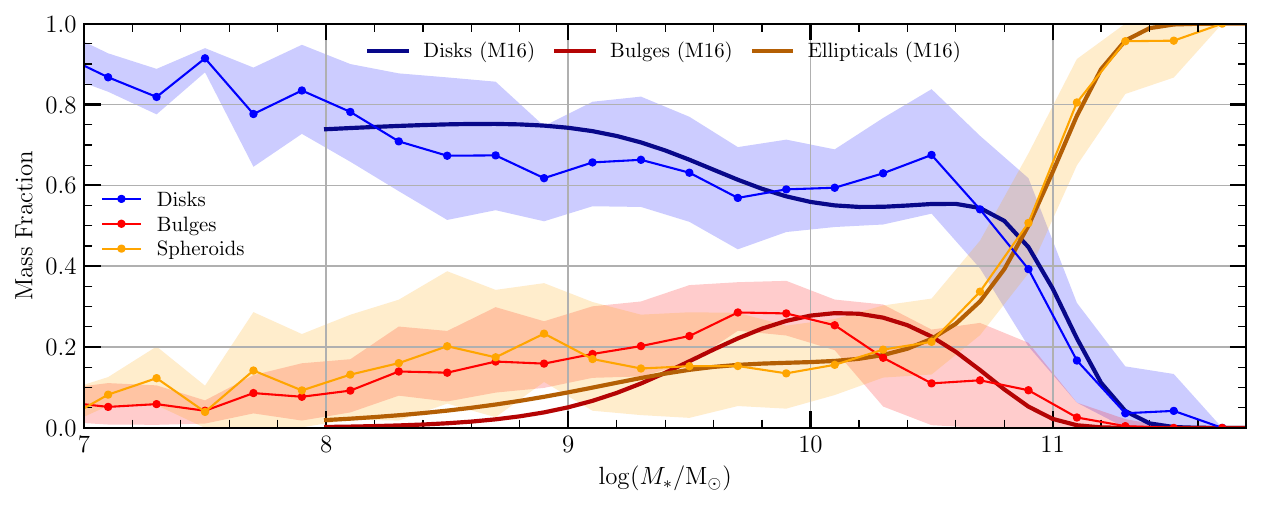}
	\caption{Fraction of total stellar mass contributed to the total as a function of component stellar mass. The equivalent curves from \citet{moffett2016b} have been included for comparison, and are shown to be entirely consistent within the expected uncertainty thanks to different classification schemes. }
	\label{fig:MassFunction_fractional}
\end{figure*}

We present the contribution of bulges, disks, and spheroids to the overall mass budget as a function of stellar mass in Fig. \ref{fig:MassFunction_fractional}, alongside the corresponding curves from \citet{moffett2016b}. 
It is remarkable that our mass fractions agree well with those measured in the past using Hubble-type classifications alone. 
At high stellar masses ($>10^{10}\rm{M}_{\odot}$), our values are indeed entirely consistent. It is just at lower stellar masses where we tend to assign more mass to both bulges and spheroids, and less mass to disks, than \citet{moffett2016b}. Note that \texttt{LBS} objects are not explicitly considered in the curves of \citet{moffett2016b} (potentially accounting for the difference in the spheroidal populations).

\subsection{Cosmic star formation history}


\begin{figure*}
	\centering
	\includegraphics[width=180mm]{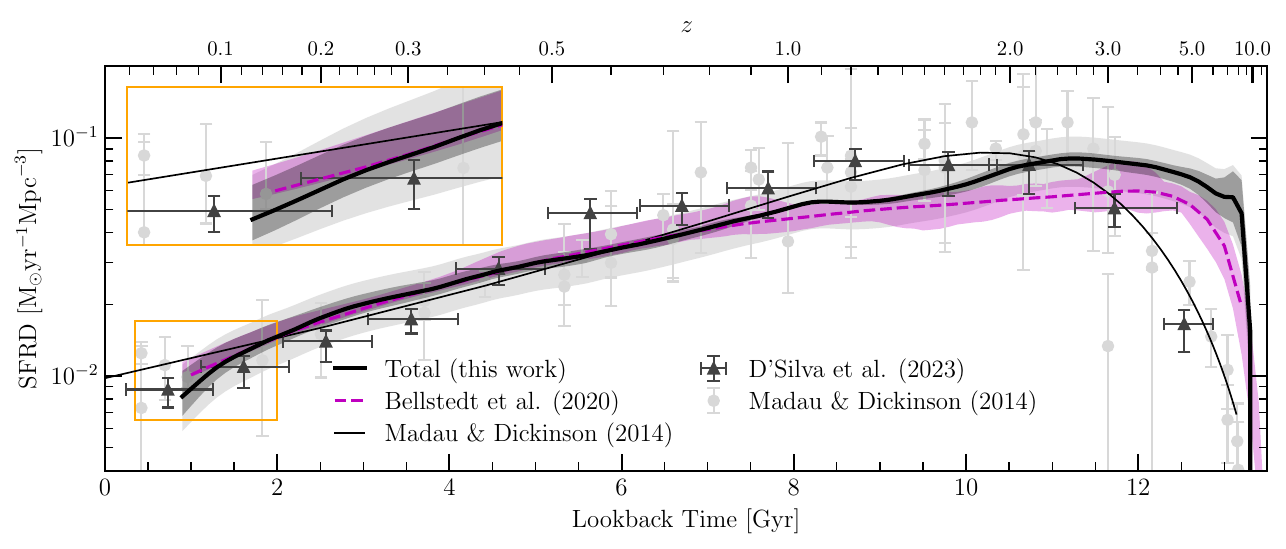}
	\caption{ CSFH derived using ProFuse in black, with classification uncertainties conveyed in dark grey. The increased uncertainty when considering cosmic variance has been shown in the lighter shaded region. The CSFH derived via ``classical" SED fitting using the code \textsc{ProSpect} from \citet{bellstedt2020b} is shown in the dashed magenta line, as are the fit to the compilation from \citet{madau2014}. The most recent observations from \citet{dsilva2023} have also been included. The inset (orange) presents a zoomed-in view of the low-$z$ portion of the CSFH. }
	\label{fig:CSFH_ProFuse}
\end{figure*}

Using our volume-limited sample, we derive the star formation rate density (SFRD) with the $z<0.06$ volume to representing the Universe. The SFRD is derived by summing the star formation histories of all the galaxies in the sample, normalised by the volume. To account for incompleteness at the lowest stellar masses (below $10^9\,\rm{M}_{\odot}$), we scale the contribution of each galaxy to the SFRD by $V/V_{\rm{max}}$,\footnote{Often referred to simply as a $1/V_{\rm{max}}$ correction} to account for the portion of the studied volume over which the galaxy is undetectable within the selection limits. 
Here, $V$ is the volume of the obervational sample, and $V_{\rm{max}}$ is the lower between the maximum volume within which the object is observable, or the full volume of the sample.

The resulting total SFRD of our sample derived using our \textsc{ProFuse} method is shown in Fig. \ref{fig:CSFH_ProFuse} as the thick black line. 
The standard compilation of data points\footnote{Measurements included in this compilation come from the following studies: \citet{sanders2003, takeuchi2003, wyder2005, schiminovich2005, dahlen2007, reddy2009, robotham2011, magnelli2011, magnelli2013, cucciati2012, bouwens2012, bouwens2012a, schenker2013, gruppioni2013}} by \citet{madau2014} is shown as faint grey data points, as well as the more recent observational measurements of the CSFH by \citet{dsilva2023}.

The forensically derived CSFH using only broadband SED fitting via \textsc{ProSpect} by \citet{bellstedt2020b}\footnote{We present here the SFRD presented in appendix B of \citet{bellstedt2020b} rather than the main body of the paper, to compare results using the equivalent metallicity evolution prescription.} is shown in Fig. \ref{fig:CSFH_ProFuse} as a dashed magenta line. 
While both this derivation and the in-situ measurement by \citet{dsilva2023} uses the modelling power of \textsc{ProSpect}, the implemented methods of deriving the SFRD are entirely different.
Although the same parametrisation for the SFH and metallicity is used within both \textsc{ProSpect} and \textsc{ProFuse}, the fact that the SFHs can effectively be multi-modal using \textsc{ProFuse} means that these CSFH derivations were by no means destined to look the same. 
Despite this, the results do agree for the most part, except for slight differences in the early Universe and small discrepancies at low redshift.

This lowest redshift range has been shown in the Fig. \ref{fig:CSFH_ProFuse} inset, highlighted in orange, so that discrepancies are clearer to see. At a lookback time of 1 Gyr, the \textsc{ProSpect} SFRD is larger than that derived by \textsc{ProFuse} by 13\%  (although the two are still entirely consistent within the uncertainty range introduced by different classification schemes). 
A possible explanation for the recent dip in the \textsc{ProFuse} CSFH is that star-forming clumps in galaxies (such as in their spiral arms) are not explicitly fitted. 
This is evident in the residuals shown in Fig. \ref{fig:RGBcompilation_bars}. 
If recent star formation is systematically under-fitted, because the structure is clumpy and not well-described by our model configurations, then we may be systematically underestimating the recent SFR in some cases. 
An additional contribution to the potential difference at recent times is the manner in which a completeness correction is conducted. A more simplistic correction was conducted by \citet{bellstedt2020b}, which may account for some differences in the lowest lookback time bins of the SFRD.

The other difference between the total CSFH in this work and that derived using \textsc{ProSpect} by \citet{bellstedt2020b} can be seen at lookback times of greater than 6 Gyr, where the added flexibility of multiple components in this work has resulted in slightly more star formation being recovered (although considering the uncertainties, the discrepancy is only small). 
Through a comparison of SED-fitting outputs from GAMA, galaxies from the semi-analytic model \textsc{Shark} \citep{lagos2018}, and SED-fitted outputs of the \textsc{Shark} galaxies, \citet{bravo2022} conducted a very thorough analysis to determine the reliability of forensic properties derived from \textsc{ProSpect}. They determined that the evolution of galaxy properties could be accurately reproduced to lookback times of $\sim$6 Gyr, but that beyond this epoch, the results start to be dominated by the modelling assumptions. Given that the light from the oldest stellar populations is much fainter than from younger ones, it is unsurprising that the constraint from SED-fitting is much lower here. For this reason, the slight deviation we see in the CSFH between the \textsc{ProSpect}-only and \textsc{ProFuse} derivations may well simply be the consequence of low constraints from these stellar populations. 
While the absolute SFHs at these older epochs likely have greater uncertainty from SED fitting that limit our derivation of absolute ages, the relative differences between histories for individual populations is still meaningful. 

The measurements by \citet{dsilva2023} are also derived using \textsc{ProSpect} with a linear metallicity evolution model, consistent with the SED modelling approach in this paper. 
\citet{dsilva2023} also include AGN in their modelling as described by \citet{thorne2022}, resulting in an improved estimation of galaxy properties, especially at high-$z$. 
Our forensic derivation is broadly consistent, however we note that our forensic SFRD is perhaps underestimated in the 8-10 Gyr range, and then overestimated beyond 10 Gyr. 
The agreement between each of the \textsc{ProSpect}-derived approaches in measuring the SFRD is remarkable within the most recent 8 Gyr of lookback time.

\begin{figure*}
	\centering
	\includegraphics[width=180mm]{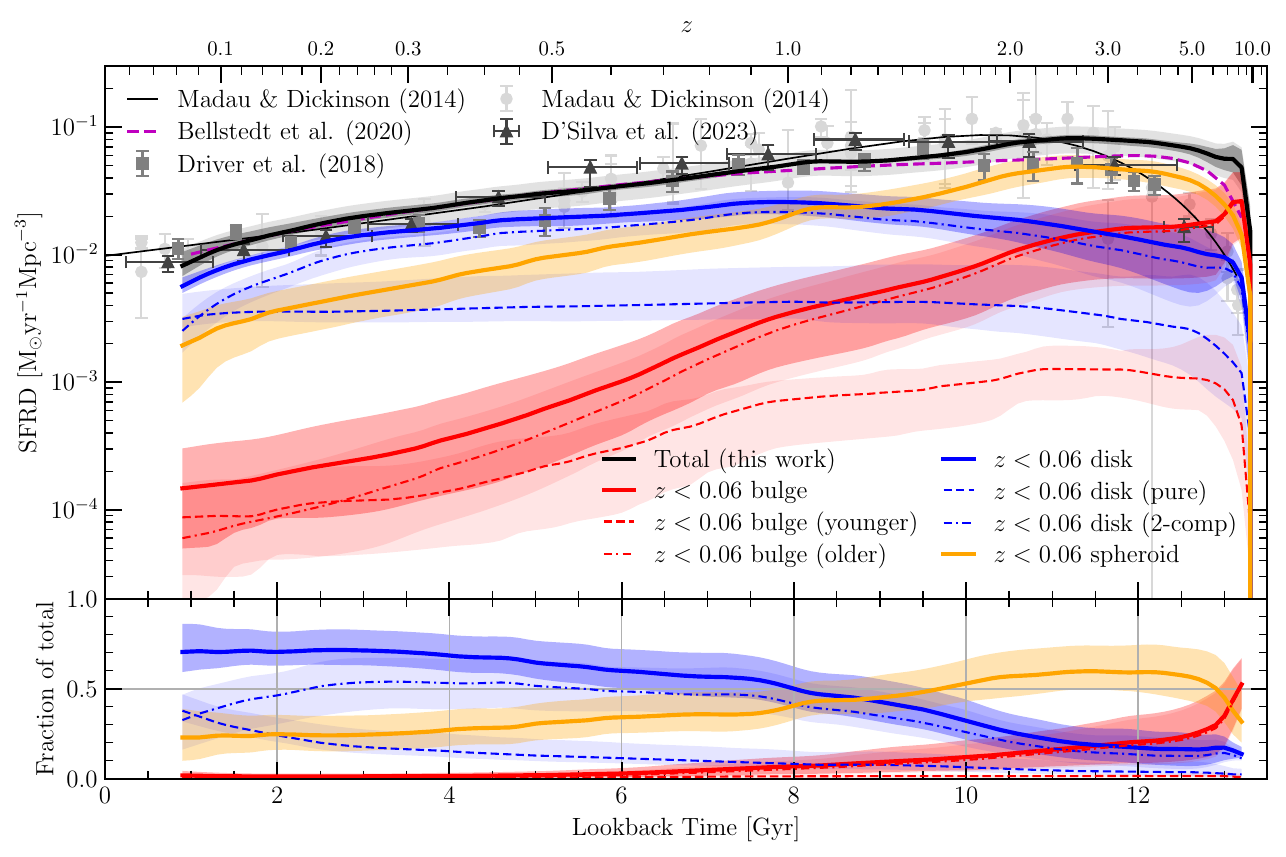}
	\caption{ CSFH as presented in Fig. \ref{fig:CSFH_ProFuse}, including the contributions from different galaxy structures. Disks are shown in blue, spheroids in orange, and bulges in red. Disks have been further divided into pure disk systems (dashed), and those from two-component systems (dashed dotted). Bulges have been divided into those that are older than their corresponding disks (dashed dotted), and those that are younger (dashed). Shaded regions indicate the uncertainty for each component that arises using different structural classification methods (auto ProFuse classifiers, visual classification for this work, and the DR4 classification by \citealt{driver2022a}).   }
	\label{fig:CSFH_ProFuse_Subcomponent}
\end{figure*}

The SFRD with contributions by different components is presented in Fig. \ref{fig:CSFH_ProFuse_Subcomponent}, with the total SFRD shown in black, and in colour the contributions toward the total SFRD by different galaxy structures, including disks (blue), bulges (red), and spheroids (orange). 
Values plotted in Fig. \ref{fig:CSFH_ProFuse_Subcomponent} are presented in Table \ref{tab:SFRDvalues}, with full data available in the supplementary material. 

\begin{table*}
	\centering
	\caption[SFRD subsets]{SFRD subsets plotted in Fig. \ref{fig:CSFH_ProFuse_Subcomponent}.   Full table available \href{https://github.com/SabineBellstedt/Bellstedt2023-SupplementaryMaterial}{online}.  }
	\label{tab:SFRDvalues}
	\renewcommand{\arraystretch}{1.5}
	\begin{tabular}{ @{}cc | cccccccc}
		\hline
		LbT & $z$ & Total &  Bulge &  Bulge (younger) &  Bulge (older) & Disk & Pure Disk & 2-comp Disk &  Spheroid \\
		Gyr & & \multicolumn{8}{c}{$\log(\rm{M}_{\odot}\rm{yr}^{-1} \rm{Mpc}^{-3})$}   \\
		\hline
		\hline

0.9 & 0.07 & $-2.09_{-0.08}^{+0.11}$ & $-3.83_{-0.47}^{+0.31}$ & $-4.06_{-0.45}^{+0.27}$ & $-4.22_{-0.51}^{+0.37}$ & $-2.25_{-0.05}^{+0.07}$ & $-2.50_{-0.07}^{+0.19}$ & $-2.60_{-0.17}^{+0.10}$ & $-2.71_{-0.45}^{+0.22}$\\
1.0 & 0.07 & $-2.05_{-0.08}^{+0.1}$ & $-3.82_{-0.48}^{+0.31}$ & $-4.06_{-0.45}^{+0.28}$ & $-4.21_{-0.51}^{+0.37}$ & $-2.21_{-0.06}^{+0.06}$ & $-2.49_{-0.07}^{+0.19}$ & $-2.54_{-0.16}^{+0.10}$ & $-2.68_{-0.43}^{+0.22}$\\
1.1 & 0.08 & $-2.02_{-0.07}^{+0.09}$ & $-3.82_{-0.48}^{+0.32}$ & $-4.05_{-0.46}^{+0.28}$ & $-4.2_{-0.51}^{+0.38}$ & $-2.18_{-0.05}^{+0.05}$ & $-2.48_{-0.07}^{+0.19}$ & $-2.48_{-0.16}^{+0.10}$ & $-2.65_{-0.39}^{+0.21}$\\
1.2 & 0.09 & $-1.99_{-0.06}^{+0.09}$ & $-3.81_{-0.48}^{+0.32}$ & $-4.05_{-0.47}^{+0.28}$ & $-4.18_{-0.51}^{+0.37}$ & $-2.15_{-0.05}^{+0.04}$ & $-2.47_{-0.07}^{+0.19}$ & $-2.43_{-0.15}^{+0.09}$ & $-2.61_{-0.35}^{+0.19}$\\
1.3 & 0.1 & $-1.96_{-0.05}^{+0.08}$ & $-3.8_{-0.47}^{+0.32}$ & $-4.05_{-0.48}^{+0.29}$ & $-4.17_{-0.46}^{+0.37}$ & $-2.12_{-0.05}^{+0.04}$ & $-2.46_{-0.07}^{+0.19}$ & $-2.38_{-0.15}^{+0.09}$ & $-2.58_{-0.31}^{+0.18}$\\
		... &  &  &  &  &  &  &  &  &   \\
		\hline
	\end{tabular}
\end{table*}

\begin{figure*}
	\centering
	\includegraphics[width=180mm]{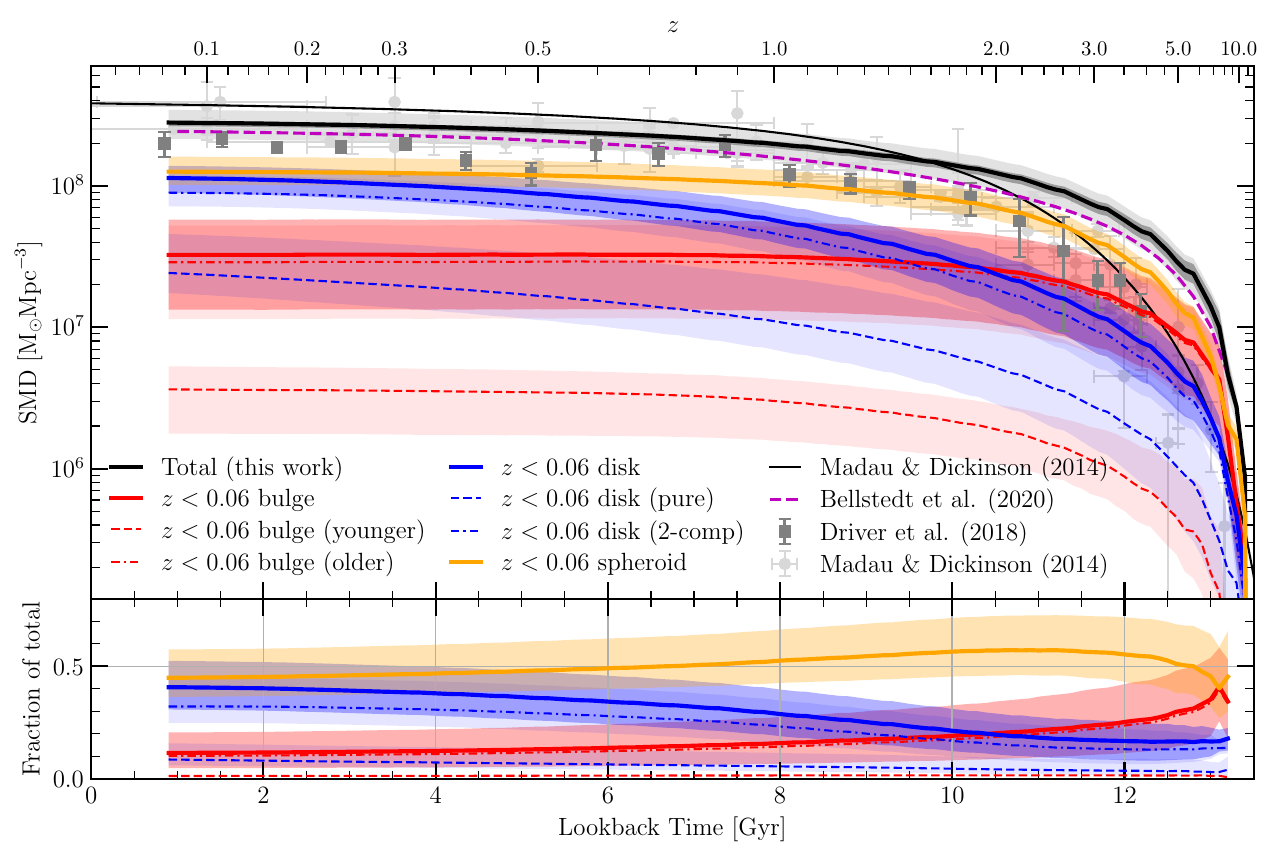}
	\caption{Stellar mass density corresponding to the SFRD presented in Fig. \ref{fig:CSFH_ProFuse_Subcomponent}, as well as the contributions by all subcomponents. }
	\label{fig:SMD_ProFuse_Subcomponent}
\end{figure*}

The corresponding evolution in the stellar mass density inferred by our star formation history is shown in Fig. \ref{fig:SMD_ProFuse_Subcomponent}. 
The impact of the slight difference in total SFRD between \textsc{ProSpect} (from \citealt{bellstedt2020b}) and \textsc{ProFuse} is a 16\% factor increase in the total stellar mass density. 
Note that despite this increase, the present-day stellar mass density is still entirely consistent with the compilation data\footnote{This compilation includes SMD measurements by \citet{arnouts2007, gallazzi2008, perez-gonzalez2008, kajisawa2009, li2009, marchesini2009, yabe2009, pozzetti2010, caputi2011, gonzalez2011, bielby2012, lee2012, reddy2012, ilbert2013, labbe2013, moustakas2013, muzzin2013}} of \citet{madau2014}. 
Values plotted in Fig. \ref{fig:SMD_ProFuse_Subcomponent} are presented in Table \ref{tab:SMDvalues}, with full data available in the supplementary material. 

We discuss the results for each component individually in the following subsections, reiterating that all trends refer to the stars that are within each component \textit{at the present day}, as opposed to the component that the stars may have been in at previous epochs.

\begin{table*}
	\centering
	\caption[SMD subsets]{SMD subsets plotted in Fig. \ref{fig:SMD_ProFuse_Subcomponent}.  Full table available \href{https://github.com/SabineBellstedt/Bellstedt2023-SupplementaryMaterial}{online}.  }
	\label{tab:SMDvalues}
	\renewcommand{\arraystretch}{1.5}
	\begin{tabular}{ @{}cc | cccccccc}
		\hline
		LbT & $z$ & Total &  Bulge &  Bulge (younger) &  Bulge (older) & Disk & Pure Disk & 2-comp Disk &  Spheroid \\
		Gyr & & \multicolumn{8}{c}{$\log(\rm{M}_{\odot}\rm{Mpc}^{-3})$}   \\
		\hline
		\hline
		
0.9 & 0.07 & $8.45_{-0.03}^{+0.01}$ & $7.51_{-0.39}^{+0.25}$ & $6.56_{-0.31}^{+0.16}$ & $7.46_{-0.40}^{+0.26}$ & $8.06_{-0.11}^{+0.08}$ & $7.38_{-0.14}^{+0.28}$ & $7.95_{-0.09}^{+0.12}$ & $8.10_{-0.09}^{+0.11}$\\
1.0 & 0.07 & $8.45_{-0.03}^{+0.01}$ & $7.51_{-0.39}^{+0.25}$ & $6.56_{-0.31}^{+0.16}$ & $7.46_{-0.40}^{+0.26}$ & $8.05_{-0.10}^{+0.09}$ & $7.38_{-0.14}^{+0.28}$ & $7.95_{-0.09}^{+0.12}$ & $8.10_{-0.09}^{+0.11}$\\
1.1 & 0.08 & $8.45_{-0.03}^{+0.01}$ & $7.51_{-0.39}^{+0.25}$ & $6.56_{-0.31}^{+0.16}$ & $7.46_{-0.40}^{+0.26}$ & $8.05_{-0.10}^{+0.09}$ & $7.38_{-0.14}^{+0.28}$ & $7.95_{-0.10}^{+0.12}$ & $8.10_{-0.09}^{+0.11}$\\
1.2 & 0.09 & $8.45_{-0.03}^{+0.01}$ & $7.51_{-0.39}^{+0.25}$ & $6.56_{-0.31}^{+0.16}$ & $7.46_{-0.40}^{+0.26}$ & $8.05_{-0.10}^{+0.09}$ & $7.38_{-0.15}^{+0.27}$ & $7.95_{-0.10}^{+0.12}$ & $8.10_{-0.09}^{+0.11}$\\
1.3 & 0.1 & $8.45_{-0.03}^{+0.01}$ & $7.51_{-0.39}^{+0.25}$ & $6.56_{-0.31}^{+0.16}$ & $7.46_{-0.40}^{+0.26}$ & $8.05_{-0.10}^{+0.09}$ & $7.37_{-0.14}^{+0.28}$ & $7.95_{-0.10}^{+0.12}$ & $8.10_{-0.09}^{+0.11}$\\
		... &  &  &  &  &  &  &  &  &   \\
		\hline
	\end{tabular}
\end{table*}

\subsubsection{Cosmic star formation history of disks}

Present-day disks are host to the vast majority of stars formed within the last 8 Gyr. 
We further divide this contribution into that from pure disks, and disks in two-component systems (shown in Fig. \ref{fig:CSFH_ProFuse_Subcomponent} and \ref{fig:SMD_ProFuse_Subcomponent} as dashed, and dashed-dotted blue lines respectively). 
While pure disk systems are responsible for more present-day star formation, this is only very recently true, due to the decline in star formation in two-component disk systems in the past 2 Gyr. 
The appearance of a recent downturn in the star formation rate density of two-component disks is unlikely to be an indication that bulges cause quenching, and likely simply the consequence of the fact that these disks are more massive, and quenched fractions are greater at higher stellar mass \citep[as discussed in many works, but see for example][]{peng2010, moustakas2013}. 

While disks do have some old stars, the fraction of stars from the early Universe ($z>2$) that resides in present-day disks is very small, at only $\sim15$\%.

\subsubsection{Cosmic star formation history of bulges}
\label{sec:bulgeCSFH}

Bulges are seen to be a very old population, with the vast majority of stellar mass in present-day bulges already formed by cosmic noon. In fact, half the stars currently in bulges had already been formed by a lookback time of 11.8 Gyrs. 
These bulge stars are distinctly older than spheroid stars.
To confirm that this is not the consequence of dust in highly inclined two-component systems, we compared the bulge CSFH for inclined and face-on systems (based on the axial ratio of the corresponding disk), seeing no bias based on galaxy inclination. 
This lends confidence that bulges truly are measurably older than spheroids. 

It is generally accepted that observationally-identified bulges encompass a range of different physical structures, including ``classical" bulges, and also ``pseudo-" bulges. These pseudobulges have distinct physical structure \citep{sandage1970, kormendy1982, kormendy2004}, and are thought to be younger than their classical counterparts \citep[for example][]{morelli2008, fisher2009}. 
It is possible that we are missing such pseudobulges in our automatic \textsc{ProFuse} classes of structure, as we do not include a model configuration that has a non-circular bulge with a free S\'{e}rsic index. 
If pseudobulge-hosting galaxies were specified as two-component objects from the visual classifications, however, then a bulge complex would have been forcibly modelled, even if it did not produce the best fit to the imaging data. 
Therefore, we expect that the stellar populations of these sources would still be represented in the visually identified bulge samples. 
Any differences between the CSFH derived by the different classification schemes would then be apparent in the uncertainty range presented in Fig. \ref{fig:CSFH_ProFuse_Subcomponent}. 
Even considering these uncertainties, bulges are still consistently substantially older than the disk population.

A small subset of the bulges in our analysis are younger than their corresponding disks (8-21\%, depending on the classification scheme\footnote{Using the \textsc{ProFuse} classifications, this value is 16\%.}), and so to specifically isolate these bulges from the rest of the population, we show in Fig. \ref{fig:CSFH_ProFuse_Subcomponent} the contribution to the bulge CSFH by ``normal" older bulges in the red dashed-dotted line, and the contribution by the relatively younger bulges in the red dashed line. 
The bulges that are relatively younger than their disks, while generally older than the disk population, actually more closely resembles the spheroid stellar populations than the typical bulges. 
A more detailed analysis of this population, and the potential implications for galaxy formation theories, are left for a future work in which this can be explored in greater depth. 

Our results indicate that either young pseudobulges are a very subdominant fraction of the bulge population (as measured by \citealt{gadotti2009} using SDSS), or pseudobulge structures have similar stellar population properties to classical bulges. 
It is also entirely possible that, because pseudobulges have properties more similar to disks than bulges in terms of bluer stellar populations and lower S\'{e}rsic indices, that they have been engulfed by the modelled disks in this analysis.  Given the number of \DR4-classified \texttt{dBD} sources (visually deemed to have a pseudobulge-like structure) that overlap with the pure disk \textsc{ProFuse} class in Fig. \ref{fig:ModelSelectionMatrix}, it is reasonable that such sources may have simply been modelled as a single disk. Futhermore it is likely that the BD model applied to pseudobulge sources will have allocated pseudobulge light to the disk component. Our overall bulge fraction may well miss the contribution of such sources.  
Nonetheless, this suggests that pseudobulges as an entity with stellar populations distinct from either classical bulges or disks are unlikely to be a dominant population. 
Further investigation is required to discriminate between these scenarios from our modelling. 
Note as well that our bulge population is quite possibly contaminated by bars (as discussed in Sec. \ref{sec:GalaxyFits}), which is capable of biasing the bulge CSFH.

\subsubsection{Cosmic star formation history of spheroids}

The more recent contribution to the CSFH by spheroids is substantially higher than that of bulges.
The epoch in which spheroid stars were dominantly formed was cosmic noon (around 9-13 Gyr ago), and while there has been a steady decline in the number of spheroid stars formed since a lookback time of $\sim$11 Gyr, spheroids are still responsible for $\sim$10-40\% of present-day star formation. 
In fact, the shape of the spheroid SFRD with time very closely resembles that of the total CSFH. 
This suggests that spheroid population is likely a mixed-bag in terms of SFH properties.

\subsection{Future improvements in CSFH measurement}

Despite the substantial effort that has been invested in recent years to better constrain the CSFH prior to cosmic noon, there is still significant uncertainty. 
This is driven by the inherent difficulty in measuring SFRs of galaxies in this epoch, due to uncertainties in the substantial dust corrections required at this epoch \citep[examples of the numerous studies that have attempted to make these measurements using both rest-frame UV and FIR include][]{kistler2009, gruppioni2013, bouwens2015,  rowan-robinson2016, novak2017, khusanova2020, khusanova2021}. 

The analysis presented in Fig. \ref{fig:CSFH_ProFuse} suggests that the capacity for codes like \textsc{ProSpect} to forensically measure the SFRD is excellent over the last $\sim$8 Gyrs. 
We therefore speculate that in future, one of the most promising methods used to constrain the $z>3$ SFRD will be to conduct a forensic analysis of galaxies at $z\sim0.8$. 
This is realistically the greatest distance at which forensic-style CSFH extractions are possible, as the stellar mass of the completeness limits becomes too high at greater lookback times, and morphologies become too irregular to model the galaxies well. 
This is also the epoch at which the uncertainties of the derived stellar mass function increase dramatically in the analysis by \citet{wright2018}. 
With this forensic approach at higher redshifts, there should then be enough constraining power within high-quality data to access cosmic dawn. 
The dataset that will make this possible is the complete spectroscopic redshift survey WAVES-deep \citep{driver2019} to be conducted on 4MOST \citep{dejong2019}, and other redshift surveys conducted with facilities like MOONS \citep{cirasuolo2020}, in combination with high-quality imaging from facilities such as Euclid \citep{laureijs2012}.

\section{Discussion}
\label{sec:Discussion}

\subsection{Literature comparison of bulge and disk formation epochs}

The overall trends presented for bulges, spheroids and disks are qualitatively similar to the parametric model presented by \citet{driver2013}, where star formation in present-day spheroids (referring in that work to all ellipsoidals) was deemed responsible for the bulk of star formation at and before cosmic noon, followed by an increase in star formation in disks. 
The subtle differences we identify are more prolonged star formation in spheroids, with a slight increase in the fraction of earlier formed stars that end up in present-day disks. 

Comparing directly with bulge/disk star formation histories from IFU-based work is difficult, because of the substantially smaller (and incomplete) samples currently studied in that level of detail. The main takeaway from an analysis of lenticular galaxies by \citet{johnston2022} is that bulges in general seem to be pretty old, but that the spread in ages for disks is much broader (and on average younger than bulges). This is consistent with the picture we see from our complete sample of $\sim$7,000 galaxies.
Even for individual subsets of galaxies however, a quantitative comparison is hampered by the fundamental differences derived in the star formation histories. For example the difficulty in distinguishing very old populations means that the star formation histories by \citet{johnston2022} begin at epochs earlier than the start of the Universe. 
Different stellar populations analysis tools, whether SED fitting tools like \textsc{ProSpect} or spectral fitting tools like  \textsc{ppxf} \citep{cappellari2004}, will all suffer from similar difficulties in constraining older populations. While there is more information in spectra than broadband photometry, there is also therefore more modelling complexity, and sensitivity to modelling degeneracies and assumptions (for example, \textsc{ppxf} is very sensitive to implementation treatments such as the use of regularization). Given the different approaches used to overcome these challenges, the quantitative histories are not yet directly comparable. 


Various observational studies have identified ``young" bulges, the existence of which seems to be at odds with our results at first glance. 
Such young bulge-like structures have been seen at the low-mass end \citep[such as the star-forming nuclear star clusters identified by ][]{johnston2020}, which would not impact the overall bulge CSFH due to the small amount of stellar mass involved. 
Young star-forming bulges were identified by \citet{barsanti2021} in an analysis of cluster galaxies. Such galaxies with bulges younger than disks do exist in our sample \citep[as shown in fig. 21 of ][and discussed in Sec. \ref{sec:bulgeCSFH}]{robotham2022}, however as this only accounts for $\sim$12\% of the two-component sample, the impact is insufficient to have an overall effect on the mass-weighted bulge CSFH. 
Nonetheless in Fig. \ref{fig:CSFH_ProFuse_Subcomponent} we separate the contribution of these younger bulges, and show that it is distinct to that of the rest of the bulge population.

Recent work by \citet{costantin2021} applied SED fitting to multi-wavelength photometric decompositions of 156 high-$z$ galaxies to study the stellar populations of bulges, disks and spheroids. 
In their work, a bimodality was identified in the formation epoch of bulges, which contributed to the interpretation by \citet{costantin2022} of two distinct phases of bulge formation in cosmic time. 
The bulge SFRD we present in Fig. \ref{fig:CSFH_ProFuse_Subcomponent} demonstrates no hint of two distinct phases. 
Instead, it suggests that bulges are simply a continuously old population, with only a small tail of star formation at recent times (contributed to by the subset of bulges that are younger than their disks). 
Even when analysing the age distribution of bulges identified in our work \citep[as presented in detail by][]{robotham2022}, we do not see any clear bimodality. 
The different redshift epochs studied hamper a direct comparison though, as our forensic approach may not be sensitive to two separate epochs that are both quite old.
If so, this suggests that any potential high-$z$ bimodality is not reflected in the properties of present-day bulges.  
Furthermore, the two studies use different SFH parametrisations (\citealt{costantin2021} use a declining delayed exponential parametrisation), which can also contribute to differences in recovered ages. 

\subsection{The ellipsoidals: spheroids and bulges}

As demonstrated by the need to clarify our nomenclature in Sec. \ref{sec:nomenclature}, the physical interpretation of the ``ellipsoidal" structures in the Universe (bulges and spheroids) varies across the literature. Are they in fact distinguishable structures with distinct evolutionary pathways (true ``eigenstructures"), or are their structural and stellar population properties similar enough to warrant interpretation as a single entity?

Our analysis with \textsc{ProFuse} consistently suggests that spheroids and bulges are distinct populations, whether that be due to distinct size--mass relations as shown by \citet{robotham2022}, the stellar mass distributions, or the overall age of the populations as shown by their distinct contributions to the CSFH\footnote{Despite taking care to indicate these CSFH resulting from a range of classification schemes, any contamination of disk-like sources into the spheroidal population could still be influencing the distinction between the bulge and spheroid CSFH. } in Fig. \ref{fig:CSFH_ProFuse_Subcomponent}. 
This conclusion is similar to that of \citet{gadotti2009}, who finds that high-mass bulges and ellipticals are offset in the size--mass relation, also concluding that they are in fact separate populations\footnote{Despite the similar conclusion, the size--mass relation for bulges measured by \citet{robotham2022} is different to \citet{gadotti2009}.}. 
It is difficult to make clear comparisons, as the spheroidal population that we analyse is a broader population than ellipticals studies by \citeauthor{gadotti2009}. 
This impact of this is seen when comparing (for example) the spheroid CSFH in Fig. \ref{fig:CSFH_ProFuse_Subcomponent} with that of the elliptical CSFH presented in fig. 8 of \citet{bellstedt2020b}, where the elliptical contribution to the CSFH dominates much earlier than the spheroids studies in this work generally. 
Our spheroids contain a larger portion of younger galaxies with more active star formation, typically at lower stellar masses.
In the past, such galaxies have been separated out into distinct classes, such as \citet{kormendy2012} who treat elliptical and dwarf elliptical (dE) galaxies as separate classes with distinct luminosity functions (see their fig. 3). 
This was also presented in the luminosity functions of \citet{binggeli1988}, where elliptical galaxies at the high-luminosity end are held distinct from dE and irregular (Irr) galaxies at the low-luminosity end. 

Contrastingly, studies of SMBH-bulge correlations treat ellipticals and bulges not as separate populations, but as a continuous population described collectively as ``spheroids" \citep[for example][]{savorgnan2016, savorgnan2016a}.
Ellipsoidals were also treated as a common-origin structure in the simple model by \citet{driver2013} (also referred to collectively as simply spheroids in that work). 

This complexity of interpretation is not clearer in simulations.  
In analysing the IllustrisTNG simulation, \citet{tacchella2019} study spheroids and bulges as one continuous class. 
This is contrasted by \citet{ortega-martinez2022}, who separate classical bulges and spheroids in their EAGLE-based analysis. Despite separating these populations into categories that seem more consistent with our observational approach, however, the spheroids in their analysis tend to be much more extended, and with a higher S\'{e}rsic index, suggesting that this divide is still not quite the same. 
In semi-analytic models like \textsc{Galform} \citep{cole2000} and \textsc{Shark} \citep{lagos2018a}, pure spheroids aren't explicitly formed. Instead, bulge-like structures are formed via two main mechanisms: mergers or disk instabilities. For this reason, ellipsoidals in these models are usually divided into disk-instability and merger-origin bulges, which are again different to any other definition of this structure. 
Yet the labels assigned to these structures are often ``pseudobulges" and ``classical spheroids" respectively \citep[as applied by][]{husko2023}, which again makes direct comparisons between studies difficult. 

The analysis from this paper, and how it relates to the literature, suggests that finding comparable definitions for disks is relatively straightforward, but that bulges/spheroids are much less clearly defined at this stage. 
This highlights that our definitions are far from universal, and it is essential that more care is taken in future to characterise the specific elements of the ``bulge complex", to facilitate more direct comparisons.

\subsection{Implications for morphological evolution}

Our forensic CSFH suggests that around 60\% of all stars formed  at $z\sim3$ end up in present-day spheroidal systems, with the other 40\% having ended up pretty evenly in both disks and bulges. 
Because our analysis is a forensic one, it simply reflects which structures stars end up in at the present day, not what type of structures they were formed in. 
Early results from JWST indicate that substantial morphological evolution has occurred since cosmic noon, with morphological classification of 850 galaxies by \citet{kartaltepe2023} suggesting that $>$56\% of galaxies at $z>3$ have a visually identifiable disk, with only 38\% of galaxies visually identified as spheroids.
To really quantify this evolution however, complete studies at high redshift are needed to estimate the mass fraction of each structure, not just the fraction of galaxies that include these structures. 

With the quality of imaging now available from JWST, it is possible to conduct bulge/disk decomposition \citep[as][did for a sample of 7 galaxies, demonstrating that a bulge component was already in place for each of them]{chen2022}.
The contrast between high-$z$ measurements and our forensic view of the early Universe clearly indicates that some morphological change must be occurring. 
The exact mechanism by which this tranformation is occurring is hard to tell from our forensic analysis. 

\begin{figure}
	\centering
	\includegraphics[width=85mm]{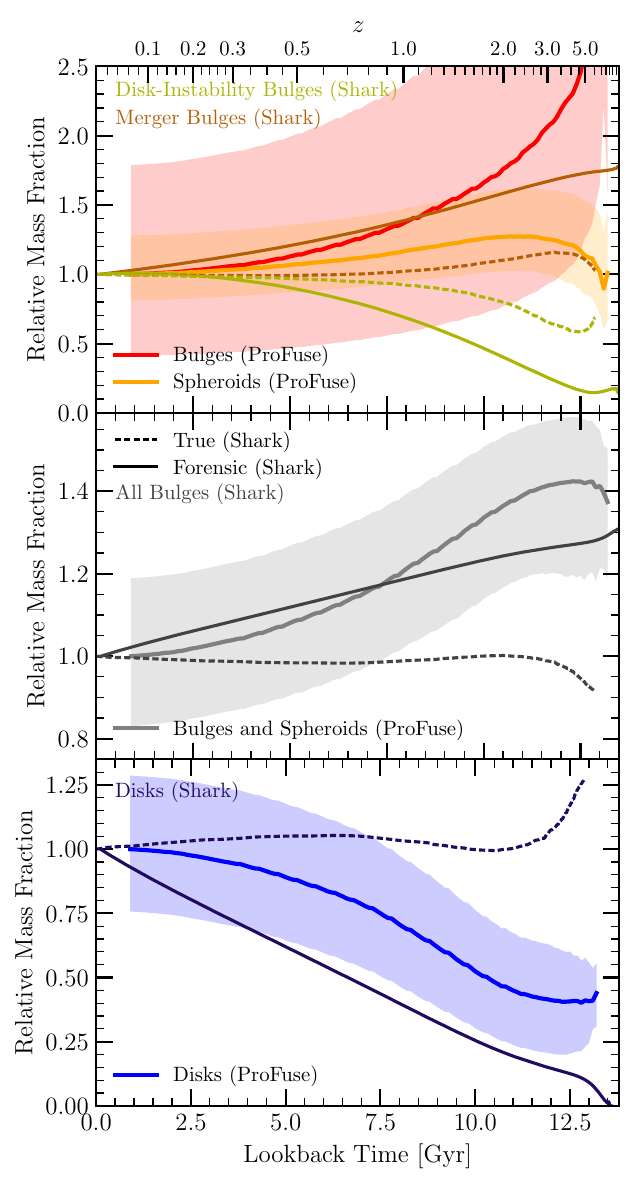}
	\caption{A comparison between the relative, fraction build-up of stellar mass in bulges and disks in our work, with the semi-analytic model \textsc{Shark}. The top panel compares bulges and spheroids by different definitions, which are not directly comparable in our work and in \textsc{Shark}. In the middle panel, however, we compare the summed contributions af all bulge/spheroidal-like structures, which are directly comparable. In the bottom panel we compare the build-up of disk mass.
	For each \textsc{Shark} contribution we have plotted two lines; the forensic view, equivalent to our analysis (solid lines), and the true view (in dashed lines). }
	\label{fig:MassFraction_withSAMs}
\end{figure}

We use the outputs from the semi-analytic model \textsc{Shark} \citep{lagos2018} to further explore the differences between ``true" mass fractions of bulges, disks and spheroids with cosmic time, as compared with forensically-derived mass fractions like the ones measured in this work. 
Seeing how these measurements differ provides some intuition for how better to interpret forensic analyses. 
This analysis is presented in Fig. \ref{fig:MassFraction_withSAMs}, where our results from the bottom panel of Fig. \ref{fig:SMD_ProFuse_Subcomponent} are renormalised to show the change in mass fraction for each structural component, relative to the $z=0$ epoch.
The absolute fractions of mass in disks and bulges is quite different in \textsc{Shark} to that seen in GAMA, which is quite possibly caused by the tuning of the semi-analytic model. For this reason we do not focus on the absolute fractions, but on the relative build-up of disk and bulge mass, which is more fundamentally linked to the modelled physics. 
The \textsc{ProFuse} results and the equivalent forensic view from \textsc{Shark} shown as solid lines. 
The ``true" view from \textsc{Shark} is shown as dashed lines. 

What the comparison of the true and forensic mass fractions from \textsc{Shark} shows, is that the historical mass in ellipsoidals is likely overestimated with a forensic approach, whereas the historical disk mass is likely underestimated. 
In the middle panel of Fig. \ref{fig:MassFraction_withSAMs}, the dark solid line shows the mass fraction in bulges assumed if \textsc{Shark} galaxies were to be analysed forensically, where the age distributions of stars in the present-day structures are analysed (as we have done observationally using \textsc{ProFuse}). 
This implies that the fraction of mass in bulges was higher at higher lookback times than in the present day (consistent with our forensic measurement). 
When the relative mass fraction is extracted from \textsc{Shark} by actually measuring this value in-situ at different epochs (shown by the dashed line), it can be seen that the fraction of stars held in bulges is actually relatively constant. 
This indicates that some of the old stars in present-day bulges must not always have been in a bulge in the past. 
Given that the reverse trend is shown in the disk component in the bottom panel of Fig. \ref{fig:MassFraction_withSAMs}, this suggests that it is possible that the old stars forensically attributed to bulges, may have actually come from disks. 
This points to a morphological evolution pathway, in which stars that are formed in disks likely end up in ellipsoidal structures.
This is consistent with the picture of morphological transformation built up based on both observations \citep[for example][]{dressler1980, hashemizadeh2022, quilley2022, marasco2023} and simulations \citep[for example][]{bournaud2007, clauwens2018, martin2018a, jackson2019}.

We caution that in different semi-analytic models, the ``true" picture can differ dramatically to that shown in \textsc{Shark}. For example, \citet{husko2023} uses \textsc{Galform} to present the fraction of mass in bulges with cosmic time, and these curves differ substantially. In particular, a much lower fraction of mass exists within bulges at the present day, but rises substantially beyond $z=2$. 

\section{Conclusions}
\label{sec:Conclusions}

We have presented here the first star formation history-based analysis to originate from an application of \textsc{ProFuse}, which simultaneously models the two-dimensional structure and SED of the components of galaxies from multiwavelength imaging. The analysis presented has the following, clear conclusions:

\begin{itemize}
	\item The capacity for \textsc{ProFuse} to produce a realistic 2D model of galaxies across multiple wavelengths simultaneously is remarkable, as demonstrated most instinctively by the colour models and residuals shown in Figs. \ref{fig:RGBcompilation_noResidual} and \ref{fig:RGBcompilation_bars} (and also \ref{fig:RGBcompilation_colourGradients}, \ref{fig:RGBcompilation_dustAsymmetry}, and \ref{fig:RGBcompilation_misc}). 
	\item Using the outputs from \textsc{ProFuse} alone through application of multiple structural configurations\footnote{We use a single component with a free S\'{e}rsic index (FS), a bulge+disk system (BD), a disk+PSF central component system (PD), or a disk+disk system (DD).}, we show that it is possible to automatically define a best-fitting model per individual galaxy. We present the differences in these best models as compared with visual classifications, and discuss the subtleties of any discrepancies. This demonstration shows that we have a pathway for moving beyond visual inspection of galaxies for the purpose of structural classification. 
	\item We showed in this paper that no two classification schemes are fully consistent, and that there are very good reasons why this is the case (most notably that the classification criteria between different schemes vary based on the purpose of the classification, be they morphological, structural, motivated by colours or not, etc). This highlights why literature results based on galaxy decompositions may differ. Rather than justifying why one method is superior to another, we took the approach of using three separate classification schemes, and using the differences to estimate uncertainty in the scientific results. We argue that this provides a fairer estimate of true uncertainties, while also lending greater confidence to the final results. 
	\item Using the star formation histories for individual galaxy components, we measure forensically the cosmic star formation history. Remarkably, the \textsc{ProFuse} CSFH is almost entirely consistent with that derived using global SED fitting of galaxies from \textsc{ProSpect} \citep{bellstedt2020b}, and also the in-situ measured CSFH from high-z galaxies \citep{madau2014, driver2018, dsilva2023}. In particular, we find that the last 8 Gyr of the CSFH are measured to be perfectly consistent across all methods. This consistency highlights not only that we have an excellent understanding of this fundamental property of our Universe, but also that the techniques we are using to study galaxies are producing consistent results, which is essential in trusting the derived properties themselves. 
	\item Finally, we present the component-wise CSFH, divided by the contributions of present-day disks, bulges, and spheroids. Bulges contain the oldest stars in the Universe, with half of all bulge stars having formed by a lookback time of 11.8 Gyrs. At the present day, a negligible fraction of all star formation occurs in bulges. Disks tell a very different story, accounting for 60-90\% of all present-day star formation, with half of disk mass in place by a much later lookback time of 7.9 Gyr. Spheroids are made of stars that have formed relatively consistently with the overall CSFH, being only slightly older than the average stars, with half the mass formed by 10.8 Gyr (as compared with an overall half-mass epoch of 9.9 Gyr). While the age estimate uncertainties of old stars from SED fitting may impact the absolute ages for our sample, we emphasise the relative difference between the populations as meaningful. Spheroids today still account for 10-40\% of all star formation, whereas bulges posess almost no star formation in the $z=0$ Universe.  
\end{itemize}


\section{Data Availability}

The data used as an input to this analysis are all publicly available via the GAMA webpage\footnote{\url{http://www.gama-survey.org}}. 
The data presented in this paper can be made available upon reasonable request.

\section{Acknowledgements}

SB and ASGR acknowledge support from the ARC Future Fellowship scheme (FT200100375). 
SPD acknowledges support from the ARC Laureate Fellowship scheme (FL220100191).
LJWD and RHWC acknowledge support from the ARC Future Fellowship scheme (FT200100055). 
This work was supported by resources provided by the \textit{Pawsey Supercomputing Centre} with funding from the Australian Government and the Government of Western Australia.

GAMA is a joint European-Australasian project based around a spectroscopic campaign using the Anglo-Australian Telescope. GAMA was funded by the STFC (UK), the ARC (Australia), the AAO, and the participating institutions. GAMA photometry is based on observations made with ESO Telescopes at the La Silla Paranal Observatory under programme ID 179.A-2004, ID 177.A-3016

We have used \textsc{R} \citep{rcoreteam2017} and \textsc{python} for our data analysis, and acknowledge the use of \textsc{Matplotlib} \citep{hunter2007} for the generation of plots in this paper. This research made use of the \textsc{R} package \textsc{celestial} \citep{robotham2016},  \textsc{Astropy},\footnote{\url{http://www.astropy.org}} a community-developed core \textsc{python} package for astronomy \citep{astropycollaboration2013, astropycollaboration2018}, \textsc{Pandas} \citep{mckinney2010}, and \textsc{NumPy} \citep{harris2020}.

\bibliographystyle{mnras}
\setlength{\bibsep}{0.0pt}
\bibliography{ZoteroLibrary}

\appendix

\section{ProFuse individual fitting outputs}
\label{appendix:FittingOutputs}

\begin{figure}
	\centering
	\includegraphics[width=85mm]{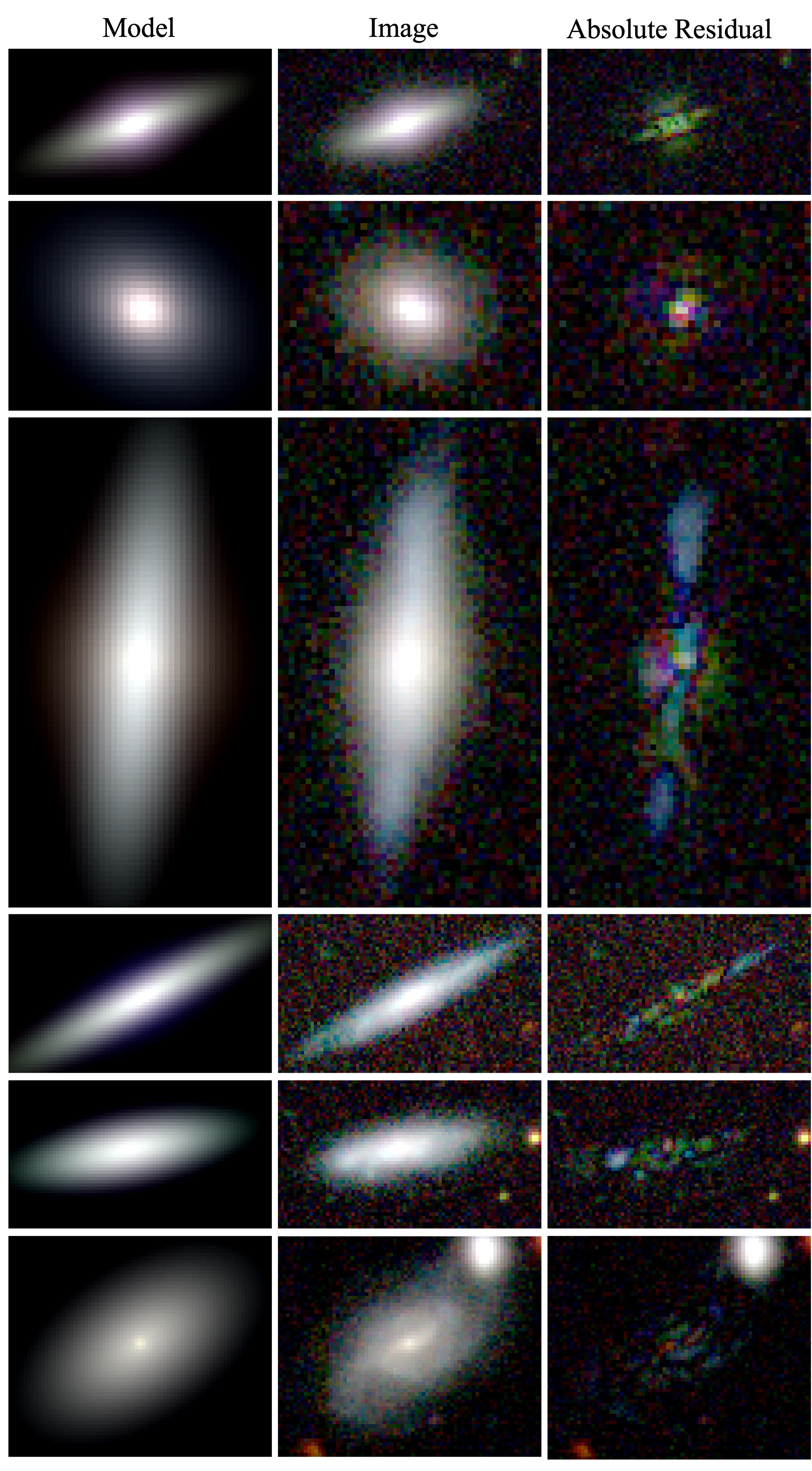}
	\caption{Compilation of RGB ProFuse ouptuts that have clear colour gradients in the model. The CATAIDs of presented galaxies are 17319, 22119, 24207, 28775, 63927, and 77713.  }
	\label{fig:RGBcompilation_colourGradients}
\end{figure}

In the main body of the text, we presented some examples of the colour images, models and residuals, for a handful of galaxies that have very little residual structure (Fig. \ref{fig:RGBcompilation_noResidual}), and also galaxies that have clear bar structures in their residual (Fig. \ref{fig:RGBcompilation_bars}).
As a further indication of the wealth of information extractable through these residual visualisations (a task that we will leave for follow-up studies), we have collated a number of examples for different features. 
Fig. \ref{fig:RGBcompilation_colourGradients} shows examples of galaxies where the different colours of the modelled structural components have created clear colour gradients in the modelled galaxy. When comparing this to the image, this seems to be sensible in most cases.  

\begin{figure}
	\centering
	\includegraphics[width=85mm]{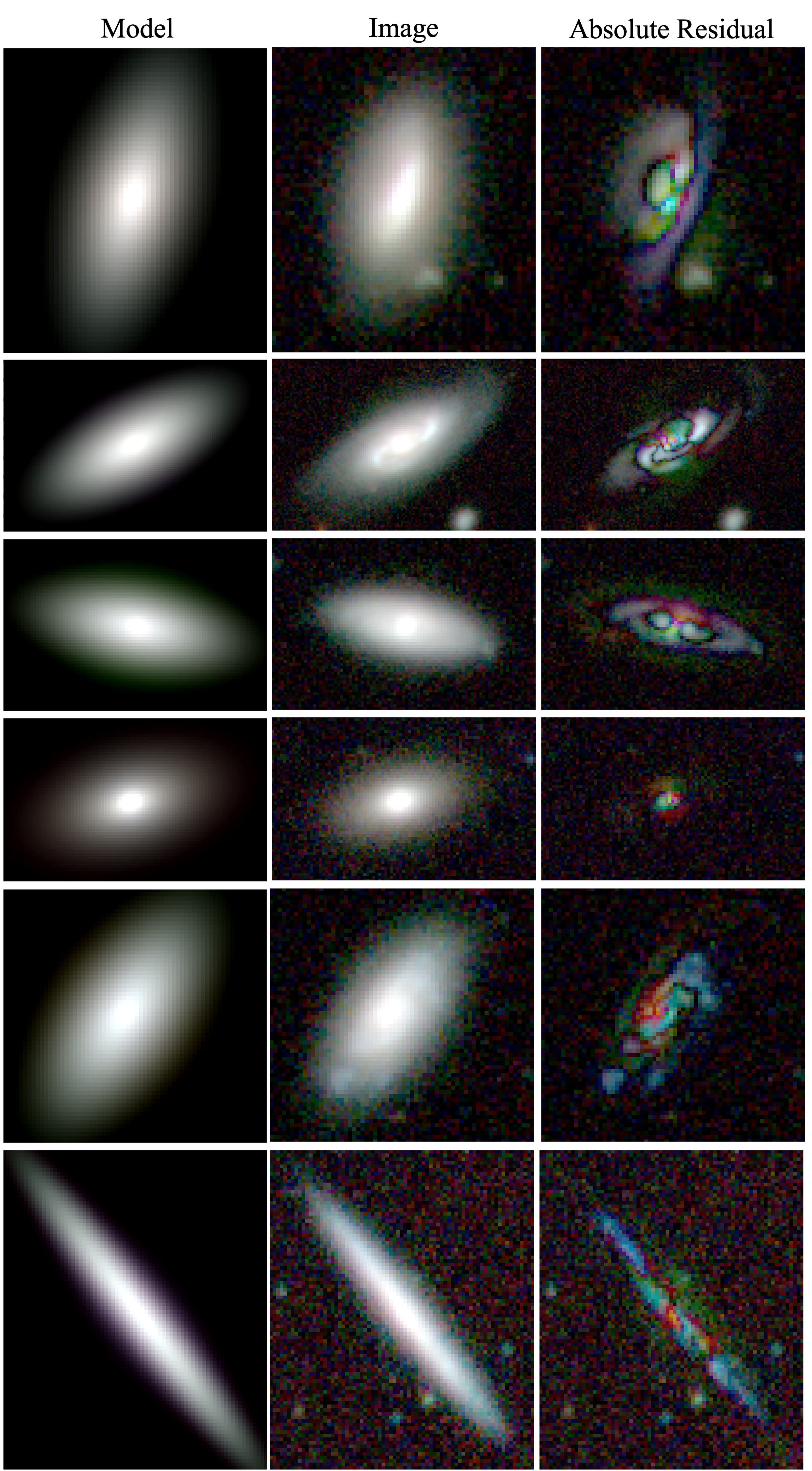}
	\caption{Compilation of RGB ProFuse ouptuts that show asymmetric features in colour, possibly indicative of dust. The CATAIDs of presented galaxies are 22805, 22834, 23075, 65416, 69983, and 77153. }
	\label{fig:RGBcompilation_dustAsymmetry}
\end{figure}

An additional type of feature identified in the examples shown in Fig. \ref{fig:RGBcompilation_dustAsymmetry}, is coloured asymmetry, usually on either side of the galaxy long axis, where one side is noticeably redder than the other. While we do not explore this feature, we suspect it is likely a signature of dust in the galaxy. 

\begin{figure}
	\centering
	\includegraphics[width=85mm]{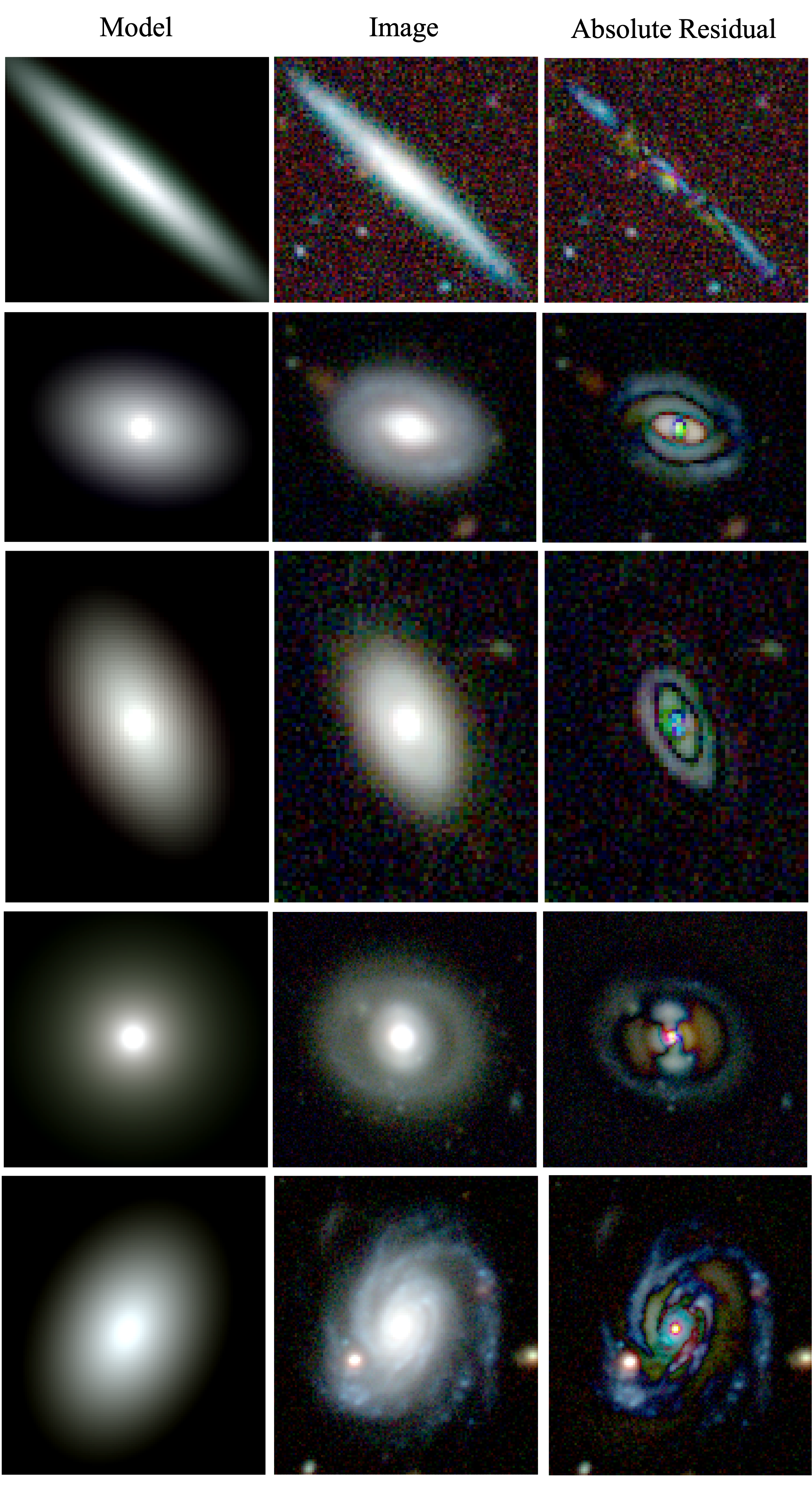}
	\caption{Compilation of RGB ProFuse ouptuts with misc features in the residual. The CATAIDs of presented galaxies are 172909, 69752, 15509, 69986, and 176955.   }
	\label{fig:RGBcompilation_misc}
\end{figure}

Finally, an assortment of miscellanous additional features is presented in Fig. \ref{fig:RGBcompilation_misc}. This includes features like very thin disks, rings, and spiral arms. These residuals are generally not an indication of a poor fit, but rather a sign that there is substructure present that has not been accounted for by the four \textsc{ProFuse} model configurations applied in this work. 

\section{Robustness of fully automated ProFuse classifications}

\begin{figure}
	\centering
	\includegraphics[width=85mm]{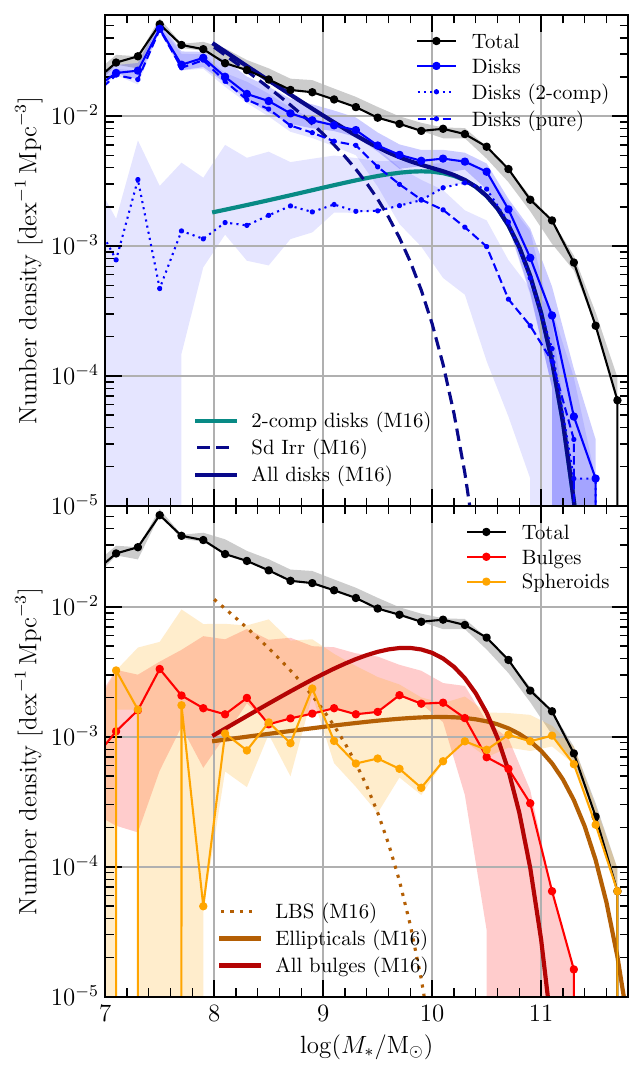}
	\caption{As Fig. \ref{fig:MassFunction_combined}, but now showing the ProFuse-only effort in solid lines.  }
	\label{fig:MassFunction_combined_bestProFuse}
\end{figure}

As a demonstration of how well the purely-\textsc{ProFuse} classifications work (as opposed to the visual classification), we show a modified version of Fig. \ref{fig:MassFunction_combined} in Fig. \ref{fig:MassFunction_combined_bestProFuse}. 

\end{document}